\documentclass[superscriptaddress]{revtex4-2}

\usepackage{fouriernc}
\usepackage{graphicx}
\graphicspath{{figures}}
\usepackage[utf8]{inputenc}
\usepackage{multirow}
\usepackage{xcolor}
\usepackage{xspace}

\definecolor{darkperiwinkle}{RGB}{102, 102, 128}

\newcommand{\orcid}[1]{\href{https://orcid.org/#1}{\textcolor[HTML]{A6CE39}{\aiOrcid}}}

\newcommand{\newacronym}[3]{%
  \newcommand{#1}{#2 (#3)\xspace%
    \renewcommand{#1}{#3\xspace}%
  }%
}

\newacronym{\GRMHD}{General Relativistic Magnetohydrodynamics}{GRMHD}
\newacronym{\MHD}{Magnetohydrodynamics}{MHD}
\newacronym{\RHS}{right hand side}{RHS}
\newacronym{\ECP}{Exascale Computing Project}{ECP}

\newcommand{\codename}[1]{\texttt{#1}}

\newcommand{\amrex}{\codename{AMReX}\xspace}
\newcommand{\cactus}{\codename{Cactus}\xspace}
\newcommand{\carpetx}{\codename{CarpetX}\xspace}

\newcommand{\theCode}{\codename{GRaM-X}\xspace}
\newcommand{\grhydro}{\codename{GRHydro}\xspace}

\newcommand{\SPEC}{\texttt{SPEC}\xspace}

\newcommand{\dF}{{^{^*}\!\!F}}
\newcommand{\Bcon}{{\mathcal B}}

\newcommand{\eqref}[1]{\eref{#1}}

\begin{document}

\title{GRaM-X: A new GPU-accelerated dynamical spacetime GRMHD code for Exascale computing with the Einstein Toolkit}

\author{Swapnil Shankar}
\email{s.shankar@uva.nl}
\homepage{\url{www.swapnilshankar.com}; orchid.org/0000-0002-5109-0929}
\affiliation{Anton Pannekoek Institute for Astronomy and GRAPPA, 
University of Amsterdam, 
Science Park 904, 1098 XH Amsterdam, The Netherlands}

\author{Philipp M\"osta}
\email{p.moesta@uva.nl}
\homepage{orcid.org/0000-0002-9371-1447}
\affiliation{GRAPPA, 
Anton Pannekoek Institute for Astronomy and Institute of
High-Energy Physics, University of Amsterdam,
Science Park 904, 1098 XH Amsterdam, The Netherlands}

\author{Steven R. Brandt}
\email{orcid.org/0000-0002-7979-2906}
\affiliation{Center for Computation \& Technology, Louisiana State University, Baton Rouge, Louisiana, USA}

\author{Roland Haas}
\email{orcid.org/0000-0003-1424-6178}
\affiliation{National Center for Supercomputing applications, University of Illinois, 1205 W Clark St, Urbana, Illinois, USA}
\affiliation{Department of Physics, University of Illinois, 1110 West Green St, Urbana, Illinois, USA}

\author{Erik Schnetter}
\email{orcid.org/0000-0002-4518-9017}
\affiliation{Perimeter Institute for Theoretical Physics, Waterloo, Ontario, Canada}
\affiliation{Department of Physics and Astronomy, University of Waterloo, Waterloo, Ontario, Canada}
\affiliation{Center for Computation \& Technology, Louisiana State University, Baton Rouge, Louisiana, USA}

\author{Yannick de Graaf}
\affiliation{University of Amsterdam, Science Park 904, 1098 XH Amsterdam, The Netherlands}

\date{2022-10-31}

\begin{abstract}
We present \theCode (\textbf{G}eneral \textbf{R}elativistic \textbf{a}ccelerated \textbf{M}agnetohydrodynamics on AMRe\textbf{X}), a new  GPU-accelerated dynamical-spacetime general relativistic magnetohydrodynamics (GRMHD) code which extends the GRMHD capability of Einstein Toolkit to GPU-based exascale systems. \theCode supports 3D adaptive mesh refinement (AMR) on GPUs via a new AMR driver for the Einstein Toolkit called \carpetx which in turn leverages \amrex, an AMR library developed for use by the United States DOE’s Exascale Computing Project (ECP). We use the Z4c formalism to evolve the equations of GR and the Valencia formulation to evolve the equations of GRMHD. \theCode supports both analytic as well as tabulated equations of state. We implement TVD and WENO reconstruction methods as well as the HLLE Riemann solver.  We test the accuracy of the code using a range of tests on static spacetime, e.g. 1D MHD shocktubes, the 2D magnetic rotor and a cylindrical explosion, as well as on dynamical spacetimes, i.e. the oscillations of a 3D TOV star. We find excellent agreement with analytic results and results of other codes reported in literature. We also perform scaling tests and find that \theCode shows a weak scaling efficiency of $\sim40$-$50\%$ on 2304 nodes (13824 NVIDIA V100 GPUs) with respect to single-node performance on OLCF's supercomputer Summit. 
\end{abstract}


\maketitle

\setlength{\parskip}{2ex}

\section{Introduction}
In the last decade dynamical-spacetime general-relativistic magnetohydrodynamics (GRMHD) codes have developed into robust tools to perform production simulations of astrophysical systems. They are routinely applied to predict gravitational waves from compact-object mergers, the amount and composition of ejected material, and to determine the remnant object left behind, e.g. ~\citep{palenzuela:13,etienne:15,kiuchi:15,paschalidis:16,ruiz:16,ciolfi:17,ciolfi:19,ciolfi:20b,moesta:20,kiuchi:22}. The outputs from these simulations are often used to identify the multimessenger signatures of these events in multi-stage pipelines~\citep{coughlin:20,raaijmakers:21}. In the supernova context, general-relativistic magnetohydrodynamics simulations have matured significantly over the last decade. Multiple groups have led proof-of-concept studies demonstrating the importance of inclusion of magnetic fields in rapidly-rotating progenitors~\citep{burrows:06,obergaulinger:09,winteler:12,moesta:14a,moesta:17a} and are beginning to study the impact of magnetic fields in neutrino-driven supernova~\citep{obergaulinger:20a}. Most of these codes employ the Valencia formulation of the ideal magnetohydrodynamics equations~\citep{anton:06} and solve them numerically via finite-volume methods while solving Einstein's equations via finite-differences or spectral methods. Much work in the last decade has gone into performing high-resolution simulations~\citep{moesta:15,kiuchi:15,Kiuchi:2015qua} and adding more realistic microphysics via tabulated equations of state and better neutrino transport approximation schemes~\citep{moesta:14a,obergaulinger:20a,kiuchi:22}. Many of these results have been enabled by making these multi-physics simulations run effectively in massively-parallel environments as found on modern high-performance computing systems.

The current challenge is to run these multi-physics simulations on modern high-performance compute systems that often contain the majority of their compute power in graphical processing units (GPUs). Standard scientific coding practices targeting employment on central processing units (CPUs) do not work effectively on GPUs due to the drastic hardware differences. Common bottlenecks that have to be taken into account are the significantly simpler control logic of GPUs, register number and capacity, memory layout and bandwidth, as well as transfer of data from CPU to GPU. Different strategies exist for programming for GPUs, namely via direct \codename{CUDA} implementation, \codename{OpenMP}, or secondary libraries like \codename{Kokkos}. A good number of \GRMHD codes that use static spacetime backgrounds have been successfully ported/redesigned to run effectively on GPUs~\citep{liska:19a,stone:20}. Dynamical spacetime \GRMHD codes however present a bigger porting challenge due to the order-of-magnitude larger memory footprint.

Here we present \theCode, a dynamical-spacetime \GRMHD code developed for the Einstein toolkit that runs efficiently on GPUs. \theCode follows the Valencia formulation of \GRMHD and can utilize analytic and tabulated equations of state. It is built on the \cactus computational framework and the \carpetx mesh-refinement driver which utilizes \amrex. It  currently utilizes a spacetime solver using the Z4c formulation of the Einstein equations. \theCode can be run on both CPUs and GPUs and uses \amrex GPU kernel launches via lambda functions. We have tested \theCode with a suite of standard \GRMHD test problems and demonstrate its performance on GPU supercomputers like ORNL's Summit on up to $\sim$ 2300 nodes.

This paper is structured as follows. In section \ref{sec:form} we briefly present the the equations solved in the code and the Valencia formulation of general-relativistic magnetohydrodynamics. We describe the numerical techniques and implementation details in section \ref{sec:impl} and present a set of code verification and sensitivity tests in section \ref{sec:tests}. We conclude by discussing the performance of the code in section \ref{sec:perf} and by summarizing and discussing future directions in section \ref{sec:summ}.

\section{GRMHD / Valencia formulation and numerical methods}
\label{sec:form}

\theCode is a dynamical spacetime \GRMHD code which means we evolve the equations of General Relativity (GR) as well as the equations of \MHD coupled together. We use the Z4c formalism~\cite{hilditch2013} to solve the equations of GR and the Valencia formulation to evolve the equations of relativistic ideal \MHD. In the ideal \MHD approximation, the fluid is assumed to have infinite conductivity and there is no charge separation.

\subsection{Z4c formulation of the Einstein equations}
The Einstein equations are a system of ten coupled second-order partial differential equations in the four-metric $g_{\mu\nu}$. We formulate the Einstein equations in the Z4c formulation~\cite{Milton:2011, hilditch2013}. This formulation is similar to the well-known BSSN formulation~\cite{Shibata:1995we, baumgarte:99, Alcubierre:2000xu, Alcubierre:2002kk}. It introduces extra dynamical fields that lead to a well-posed formulation and which dynamically dampens (makes decay) the constraints of the Einstein equations. The gauge conditions associated with this formulation are the $1+\log$ foliation and $\Gamma$-driver shift. We use standard gauge and constraint damping parameters for our calculations which include constraint damping parameters $\kappa_1 = 0.02$ and $\kappa_2 = 0.0$, as well as lapse parameter $\mu_L = 2/\alpha$ and shift parameters $\mu_S = 1$ and $\eta = 2$.

As usual in the Einstein Toolkit, the Z4c state vector is not exposed to other thorns in Cactus. Instead, other thorns are written in terms of the standard ADM variables: the 3-metric $\gamma_{ij}$, the extrinsic curvature $K_{ij}$, lapse $\alpha$, shift $\beta^i$, the time derivative of the lapse $A=\partial_t\alpha$, and the time derivative of the shift $B^i=\partial_t\beta^i$.

\subsection{Valencia formulation}\label{valencia_formulation}
The equations of ideal \GRMHD used in \theCode are obtained from the conservation of mass and energy-momentum as well as from the Maxwell's equations:
\begin{equation}
  \nabla_{\!\mu} J^\mu = 0\,\,, \qquad \nabla_{\!\mu} T^{\mu \nu} = 0\,\,, \qquad \nabla_\nu \dF^{\mu\nu} = 0\,\,
  \label{eq:equations_of_grmhd_origin}
\end{equation}
where $ \nabla_{\!\mu} $ denotes the covariant derivative with respect
to the 4-metric, $J^{\,\mu} = \rho u^{\,\mu} $ is the mass current and $ \dF^{\mu\nu}$ is the dual of the relativistic Faraday tensor $F^{\mu\nu}$.  In the ideal \MHD approximation, electric fields vanish in the rest frame of the fluid which leads to the condition:
\begin{equation}
E_\nu = u_{\mu} F^{\mu\nu}=0\,\,
\end{equation}

The equations of \MHD are coupled to the equations of GR via the stress-energy tensor $T^{\mu \nu}$. The stress-energy tensor has both the hydrodynamic contribution $T^{\mu \nu}_{\rm H}$ and the electromagnetic contribution $T^{\mu \nu}_{\rm EM}$ given by:
\begin{eqnarray}
T^{\mu \nu}_{\rm H} &=& \rho h u^\mu u^\nu + P g^{\mu \nu}  = \left( \rho + \rho \epsilon + P \right) u^\mu u^\nu + P g^{\mu \nu}    \label{hydro-stress-tensor} \\
T^{\mu \nu}_{\rm EM} &=& F^{\mu \lambda} {F^{\nu}}_\lambda  - \frac{1}{4} g^{\mu \nu} F^{\lambda \kappa} F_{\lambda \kappa} 
= b^2 u^\mu u^\nu - b^\mu b^\nu  + \frac{b^2}{2} g^{\mu \nu}  \label{em-stress-tensor} 
\end{eqnarray}
where $\rho$, $\epsilon$, $P$, $u^\mu$, and $h\equiv
1+\epsilon+P/\rho$ are the fluid rest mass density, specific
internal energy, gas pressure, 4-velocity, and specific enthalpy, respectively, and
$b^\mu  = u_\nu \dF^{\mu \nu}$ is the magnetic $4$-vector (the projected component of the
Maxwell tensor parallel  to the 4-velocity of the fluid). The combined stress-energy tensor is given by:
\begin{eqnarray}
T^{\mu \nu} 
  &=&  \left( \rho +  \rho \epsilon + P + b^2 \right) u^\mu u^\nu + 
       \left(P + \frac{b^2}{2} \right) g^{\mu \nu} - b^\mu b^\nu \label{mhd-stress-energy-tensor} 
  \\
  &\equiv& \rho h^*u^\mu u^\nu + P^* g^{\mu \nu} - 
           b^\mu b^\nu, \nonumber  
\end{eqnarray}
where $P^*=P+b^2/{2}$ is the fluid pressure combined with magnetic pressure, and $h^*\equiv 1+\epsilon+\left(P + b^2\right)/\rho$.
In \theCode, fluid variables are stored as cell averages while the spacetime variables and the stress-energy tensor are stored as samples at the cell vertices. To calculate $T^{\mu \nu}$ in Eq. \ref{mhd-stress-energy-tensor}, we need all the variables at cell vertices. Hence, we perform a 4th-order symmetric 3D interpolation~\cite{albin:hal-00967383} from cell center values to calculate the fluid variables at cell vertices.  

The evolution equations of ideal relativistic \MHD that we use in \theCode are written in a first-order hyperbolic flux-conservative form for the conserved variables $D$, $S^i$, $\tau$, and $\Bcon^i$. The conserved variables are related to the primitive variables $\rho$, $\epsilon$, $v^i$, and $B^i$ as
\begin{eqnarray}
  D &=& \sqrt{\gamma} \rho W\,\,,\label{eq:p2c1}\\
  S_j &=& \sqrt{\gamma} \left(\rho h^* W^{\,2} v_j-\alpha b^0b_j\right)\,\,,\label{eq:p2c2}\\
  \tau &=& \sqrt{\gamma} \left(\rho h^* W^2 - P^*-(\alpha b^0)^2\right) - D\label{eq:p2c3}\,\,, \\
\Bcon^k&=&\sqrt{\gamma}B^k\,\,,\label{eq:p2c4}
\end{eqnarray}
where $B^i = n_{\nu} \dF^{i\nu}$ is the spatial magnetic field in the spacelike slice with unit normal $n^\mu$, $ \gamma $ is the determinant of $\gamma_{ij} $, $W \equiv (1-v^i v_i)^{-1/2}$ is the Lorentz factor and $v^i$ is the 3-velocity defined as
\begin{equation}
v^i = \frac{u^i}{W} + \frac{\beta^i}{\alpha}\,\,,
\label{eq:vel}
\end{equation}

The \MHD evolution equations, also known as Valencia formulation, are
\begin{equation}
  \frac{\partial \mathbf{U}}{\partial t} +
  \frac{\partial \mathbf{F}^{\,i}}{\partial x^{\,i}} =
  \mathbf{S}\,\,,
  \label{eq:conservation_equations_gr}
\end{equation}
with
\begin{eqnarray}
  \mathbf{U}  ~=~ & &[D, S_j, \tau,\Bcon^k]\,\,, \nonumber\\
  \mathbf{F}^{\,i} ~ = ~&& \alpha\times
  \left[\begin{array}{c} 
          D \tilde{v}^{\,i}\\ 
          S_j \tilde{v}^{\,i} + \sqrt{\gamma} P^* \delta^{\,i}_j - b_j\Bcon^i/W\\
          \tau \tilde{v}^{\,i} + \sqrt{\gamma} P^* v^{i}-\alpha b^0 \Bcon^i/W\\
          \Bcon^k\tilde{v}^i-\Bcon^i\tilde{v}^k 
        \end{array} \right]\,\,,   \label{eq:definition-of-flux} \\
  \mathbf{S} ~ =~ && \alpha \sqrt{\gamma} \times\left[\begin{array}{c}
         0, \\ 
         T^{\mu \nu} \left( \frac{\partial g_{\nu j}}{\partial x^{\,\mu}} - 
                      \Gamma^{\,\lambda}_{\mu \nu} g_{\lambda j} \right) \\
         \qquad\alpha \left( T^{\mu 0}
         \frac{\partial \ln \alpha}{\partial x^{\,\mu}} -
         T^{\mu \nu} \Gamma^{\,0}_{\mu \nu} \right)\\
         \vec{0} \end{array}\right]\,\,.
  \label{eq:definition-of-source}
\end{eqnarray}%
Here, $ \tilde{v}^{\,i} = v^{\,i} - \beta^i / \alpha $ and $
\Gamma^{\,\lambda}_{\mu \nu} $ are the 4-Christoffel symbols. \\  

\subsection{Numerical methods}

As described above, the Valencia formulation for \GRMHD consists of a set of 8 coupled hyperbolic partial differential equations for the 8-element state vector $\mathbf{U}$. The state vector contains only the so-called \emph{conserved} variables. The so-called \emph{primitive} variables need to be calculated from the conserved variables before the fluxes $\mathbf{F}$ can be calculated. The primitive variable vector is $\mathbf{P} = [\rho, v^i, \epsilon, B^i]$.

Our initialization scheme proceeds as follows:
\begin{enumerate}
\item Initial conditions are set up in terms of the primitive variables $\mathbf{P}$ (and the ADM variables for the spacetime metric).
\item From these, the conserved variables $\mathbf{U}$ are calculated. This step is straightforward.
\end{enumerate}

Our evolution scheme is based on the \emph{Method of Lines}, allowing us to use a common time integration mechanism for all evolved variables, i.e. for both spacetime and hydrodynamics quantities. In the method of lines, one needs to provide a so-called \RHS function that calculates $\partial_t\mathbf{U}$ from a given $\mathbf{U}$. The time integrator evaluates the \RHS function multiple times to step from a solution at time $t$ to a solution at time $t+\Delta t$.

Evaluating the \RHS proceeds as follows, starting from the state vector $\mathbf{U}$ and ending with its time derivative $\partial_t\mathbf{U}$:
\begin{enumerate}
\item Find the primitive variable vector $\mathbf{P}$ from the conserved variable vector $\mathbf{U}$. This step is non-trivial because it requires finding the root of a multi-dimensional nonlinear equation for each grid cell. Luckily, the cells are not coupled, so that this expensive step can be easily parallelized. This step also requires evaluating the EoS (equation of state---see below), which might require interpolation in a nuclear EoS table. This is the most complex step of our scheme, and it can fail in several ways. The scheme can fail to converge or converge to unphysical values such as negative density, negative velocity, velocity greater than the speed of light, or negative pressure. Moreover, even if it converges to physical values, the obtained or intermediate values at any step can go out of bounds from tabulated values of $\rho$, $T$, $Y_e$ or any other dependent quantity in the table. 
\item The primitive variables are now known at the cell centers. We reconstruct their values on the cell faces using either a TVD (total variation diminishing)~\cite{toro:99} or a WENO5 (5\textsuperscript{th} order weighted-ENO)~\cite{shu:98} scheme. This step also requires the spacetime metric, which is known at the cell vertices and is interpolated linearly to the face centers.
\item For a given cell face, reconstruction from left and right side lead to potentially discontinuous hydrodynamic states on either side of the interface. We construct these Riemann problems at all the interfaces and solve them using Riemann solvers. We use an HLLE
(Harten-Lax-van Leer-Einfeldt)~\cite{Einfeldt:1988og, Harten:1983hr, Gammie:2003rj} solver.
\item The solutions of these Riemann problems provide us with the net flux across all the faces of a given cell. The divergence of this flux defines the flux term of the \RHS (\ref{eq:definition-of-flux}) together with the source term (\ref{eq:definition-of-source}). This completes the \RHS evaluation.
\end{enumerate}

We have implemented TVD and WENO reconstruction methods in \theCode. TVD is 2\textsuperscript{nd}-order accurate in regions of smooth, monotonic flows but reduces to first order in the presence of extrema and shocks. For WENO, we have implemented a version which is 5\textsuperscript{th} order accurate for smooth, monotonic flows~\cite{shu:98} and this is the reconstruction method we plan to use in our production simulations. The reconstruction method can be set using a runtime parameter.

We employ an HLLE Riemann solver in \theCode. This is an approximate solver which is less expensive numerically. More complex solvers such as Roe and Marquina have been numerically very resource intensive on traditional CPU-based codes, but we plan to implement them in the future in \theCode to extract the full compute capability of GPUs while attaining higher accuracy. This is because methods which are compute intensive but not memory intensive add little extra cost on GPUs.

Another important aspect of a \GRMHD code is the equation of state (EoS). We employ analytic equations of state such as Polytropic and Ideal Gas EoS, as well as realistic nuclear equations of state made available in the form of tables. For a Polytropic equation of state, fluid pressure is given by $P = K \rho^{\gamma} $, where $K$ is the Polytropic constant and $\gamma$ is the Polytropic index.  For an ideal gas  (also known as $\Gamma$-law ) equation of state, the fluid pressure is given by $P = (\Gamma-1)\rho\epsilon$. For nuclear tabulated equations of state, a total of 19 fluid variables such as pressure $P$, specific enthalpy $h$, the speed of sound $c_s$ etc. are given as a function of ($\rho, T, Y_e$) in the form of a table, where $T$ is the temperature and $Y_e$ is the electron fraction~\cite{Buyukcizmeci_2014}.

To find the value of a given variable in a table, one needs to look up those values in the table and perform any necessary interpolations. Since we have a GPU-based code, we load the table in the \texttt{unified memory} which is accessible from both the CPU (\emph{host}) and the GPU (\emph{device}). 
We use tables in the format as found on~\cite{stellarweb}.

The recovery of primitive variables $\rho$, $v^i$, $\epsilon$, and $B^i$ from the conserved variables $D$, $S^i$, $\tau$ and $\mathcal{B}^i$ is in principle a five-dimensional (5D) root-finding problem involving the inversion of Eqs.~\ref{eq:p2c1}-\ref{eq:p2c3}, given that inverting the magnetic field components is trivial because they just differ by a factor of $\sqrt{\gamma}$. This, in turn, renders the unknown, first term of $S^i$ in Eq.~\ref{eq:p2c2} collinear to $v^i$, thus reducing the dimensionality of the recovery problem to 3D.

In \theCode, we use either of two methods to perform this conservative to primitive transformation: A 3D Newton-Raphson (3D-NR hereafter)~\cite{Siegel_2018, cerda_duran:2008} root finder and the method of Newman \& Hamlin (Newman's method hereafter)~\cite{Newman_Hamlin:2014, Siegel_2018}. In 3D-NR, the system of equations is converted to 3 equations in unknowns ($W$, $z=\rho h W^2$, $T$) and solved for them, as described in~\cite{Siegel_2018}. Newman's method is an effective 1D method in which we iterate over the fluid pressure to find a solution. It solves a cubic polynomial of the form $f(\epsilon)=\epsilon^3 + a\epsilon^2+d$ in variable $\epsilon \equiv B^2 + z$, where $a$ and $d$ are determined by fluid variables. This method requires the calculation of $P$ from $\rho$ and $h$ using the EoS at every iteration step.

In practice, we use 3D Newton-Raphson as the primary method for \texttt{con2prim}. If 3D-NR does not converge, we fall back to Newman's method. The reason for choosing 3D-NR as the primary method over Newman's method is that the number of EoS calls is $\sim 20$ times larger in case of Newman's method compared to 3D-NR~\cite{Siegel_2018}, which makes Newman's method more expensive computationally. If Newman's method does not converge either, then we use bisection in temperature  because it is guaranteed to converge to a solution, but is much more expensive computationally. 

\section{Implementation}
\label{sec:impl}
\theCode is a complete redesign of \grhydro~\cite{moesta:14a} which we have used successfully in production for many years~\cite{moesta:14a, moesta:15a, moesta:17a, halevi:18a, moesta:20, curtis:21}. Like \grhydro, \theCode is based on the \cactus computational framework~\cite{cactus:web, CactusUsersGuide:web}, and uses the
upcoming \carpetx driver to provide mesh refined grids, I/O and inter-node communication. The \cactus framework has been in use since 1998~\cite{Bona:1998dp} and has been chosen twice as a \SPEC benchmark module and has been used in hundreds of scientific publications.

\cactus provides a ``flesh'', which works as a connection layer between end-user provided application code,
``thorns''. Most of the functionalities necessary for complex multi-physics simulations are provided by the thorns. These thorns use \cactus' domain-specific
language (DSL) to schedule subroutines, define interfaces for externally accessible subroutines, etc. All the desired thorns are given in a thornlist at compile time, but runtime parameters select which of these thorns are active in a given simulation.

\begin{figure}[htbp!]
\centering
\includegraphics[width=0.7\linewidth]{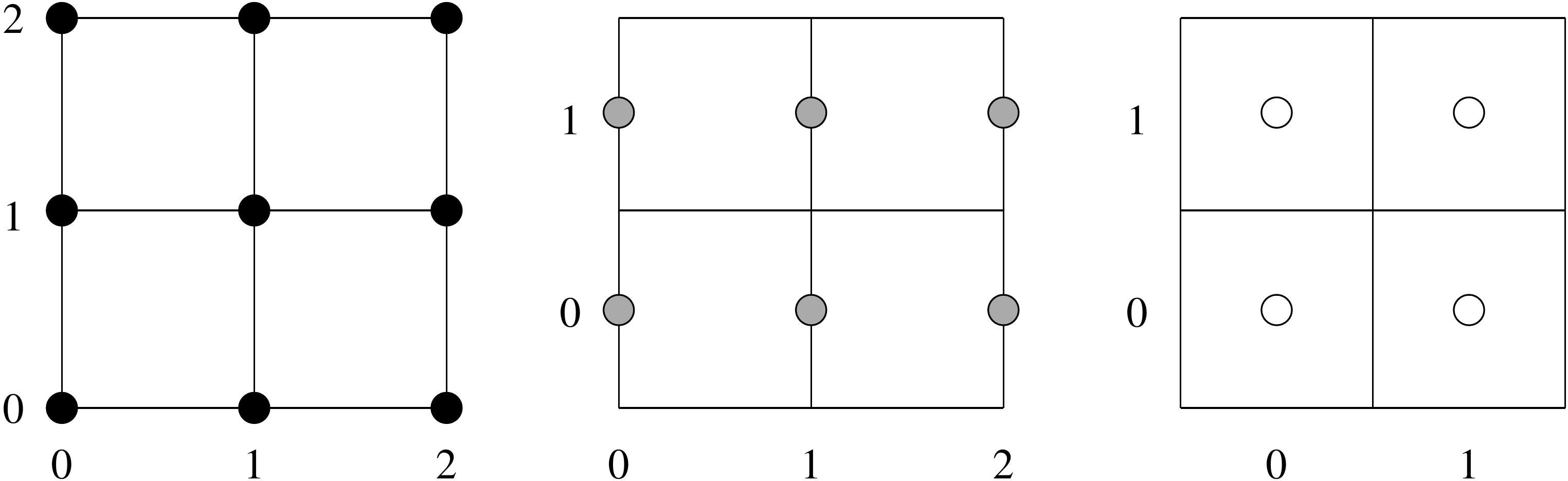}
\caption{A sample of data arrangements supported by \carpetx, displaying
different types of variables. From left to right: fully vertex centered, face
centered in the $y$ direction, and fully cell centered. Notice that there is
an extra value in
any vertex centered direction compared to cell centered
directions. This figure was modeled after~\cite{amrex-docs:web}.
}\label{fig:amrexindextypes}
\end{figure}

Of preeminent importance for the performance of a simulation is the ``driver'' thorn. This special thorn handles all data movement and memory allocations, it implements the simulation workflow, and provides I/O services. Our driver is \carpetx. It implements Berger-Oliger block-structured AMR, calling user defined functions on blocks of data for refined sections of grid as needed. \carpetx also provides parallel I/O using the \codename{Silo} and \codename{openPMD} libraries, high-order prolongation and restriction operators for variables defined at different grid locations (see figure~\ref{fig:amrexindextypes} for an illustration), inter-process exchange of ghost-halo data using the \codename{MPI-2} standard, intra-process parallelization using \codename{OpenMP}, and collective data reduction operators for computing norms of grid functions.

We have developed \carpetx over the course of the past 3 years. \carpetx leverages the
\amrex adaptive mesh refinement library that is developed for use by the DOE's \ECP~\cite{ExascaleComputingProject_:web}.
With \amrex, our codes leverage the support of the \amrex
community to ensure that the underlying AMR driver scales to ever larger systems and new architectures.
\amrex already implements algorithms to enable efficient use of current CPUs, and GPUs, as well as asynchronous iterators overlapping computation with communication, which benefit all our codes.  The continued support for \amrex through the
\ECP co-design approach~\cite{ecpamrexcodesign_:web}, where it is currently used by 7 applications funded by \ECP, provides a stable basis for \carpetx.

We have extended the inter-mesh transport operators in \amrex to support higher order accurate operations based on generic templated stencils to support the vertex centered high accuracy
operations used for spacetime geometry quantities as well as cell centered conservative operations used for fluid quantities.
All of \carpetx's basic operations are fully GPU enabled using
\amrex's GPU facilities, thus avoiding costly data transfers from GPU to CPU memory. This includes facilities to extract gravitational waves.

Spacetime geometry quantities are stored on vertices, ensuring high accuracy in inter-mesh interpolation operations, while fluid quantities are stored as cell averages in cells and updated in a flux-conservative scheme ensuring conservation of rest mass in a finite volume scheme. We pre-compute derivatives and fluxes using temporary storage locations and improve cache utilization (on CPUs).

When executing on CPUs, we use the \codename{NSIMD} library~\cite{NSIMD:web} to achieve efficient SIMD vectorization of arithmetic operations. We also use a set of classes to handle vector and tensor objects and their interactions, such as dot products and small matrix inversions. These classes can also be used with NSIMD. We use \codename{OpenMP} and grid tiling methods to distribute CPU work among
multiple cores on a single compute node while ensuring efficient use of L2 and L3 caches. For GPUs \amrex supports \codename{CUDA}, \codename{HIP/ROC} or \codename{DPC++/SYCL} as applicable for accelerator access.

Inter-node parallelism is handled using \codename{MPI} as provided by \amrex. Cells at inter-processes and inter-mesh interfaces are filled via ghost zone (halo) exchange and prolongation (interpolation) from coarse to fine grid. At
the same time, at the outer boundaries of the simulation domain, either periodic or a user-supplied boundary condition are applied.

\carpetx uses \codename{OpenPMD} and its \codename{ADIOS2} backend to output checkpoint data, which we found to be more efficient than \amrex's built in \codename{BoxLib} output. We employ \codename{Silo} for 3D grid data output, which can be directly read by \codename{VisIt} for visualization using visualization resources available at compute centers.

Our code contains extensive facilities to enforce correctness of the data access, tracking which parts of the grid are valid and detecting access attempts to invalid or out of date data. These facilities can be enabled via runtime parameters and are typically enabled for development and test simulations but disabled for production runs due to their impact on performance.

%
%

\subsection{Adaptive Mesh Refinement}

\amrex (and thus also \carpetx) provides so-called \emph{block-structured mesh refinement}. This means that there is a rectangular \emph{coarse grid}, and this coarse grid is overlaid by various \emph{refined grids}, which are also rectangular, and which are organized in \emph{refinement levels}. Each level has half the spacing of the next coarser level. The refined grid must be \emph{properly nested}, which means that the grid of level $L+1$ must be wholly contained in the grids of level $L$. 

During \emph{regridding}, one needs to flag which cells of which grid need to be refined, and which currently refined cells are not needed any more. \amrex will then combine the region of flagged cells into new, rectangular grids, enlarging the refined region somewhat if necessary. This ensures that the new refined region can easily be described by a set of rectangles. One assumes that the cost of the AMR algorithm scales not only with the total number of refined cells, but also with the number of rectangular grids, so that having fewer (but slightly larger) grids can be desirable. 

Each refined grid is surrounded by a layer of \emph{ghost zones}. These ghost zones are defined via interpolation (\emph{prolongation}) from the next coarser level. Altogether, this means that the evolution system need only be implemented for rectangular arrays, and does not need to be immediately aware of the shape and relation of the refined regions. It also need only handle a single grid resolution, since sufficient ghost zones are provided by the AMR algorithm. 

Finally, since the refined regions overlay coarser regions, this leads to a double covering of the domain. It is necessary to keep these consistent. When a refined grid's ghost zones are filled via prolongation from the next coarser grid, the coarse grid regions that are overlaid by the refined grid is reset to the respective fine grid values (\emph{restriction}). 

During time evolution, the CFL (Courant-Friedrichs-Lewy) criterion states that coarser grids can take larger time steps than finer grids. This is also called \emph{subcycling in time}, since the finer grids take more time steps than the coarser grids. For simplicity, we do not implement this yet; instead, we use the same time step size for all refinement levels. This increases the computational cost somewhat, but also increases parallel scalability (see section \ref{sec:perf} below.), since all refinement levels can be evolved simultaneously. Subcycling in time necessarily serializes evolving different refinement levels. 

\amrex's AMR algorithm structures the refined grids as rectangular blocks of a given size that can be chosen at runtime. A typical block size would be $8\times8\times8$ cells. This means each refined grid is always a multiple of e.g. $8$ cells large in each direction. 

\subsection{Staggered Grids}

It is often convenient to stagger evolved variables with respect to each other. For example, fluxes (section \ref{valencia_formulation}) are naturally defined on cell interfaces located halfway in between cell centers (\emph{staggered}). \amrex allows quantities to be located either at cell centers, on cell faces, cell edges, or at the cell vertices. Its refinement scheme is based on cells. Quantities living on faces, edges, or vertices then live on grids that are one point larger in certain directions~\cite{amrex-docs-homepage:web}.

Being able to stagger fluxes between evolved conserved quantities, or placing the magnetic vector potential $A^i$ at cell edges, has important advantages, in particular near mesh refinement interfaces where coarser and finer grids meet. For example, such schemes allow quantities to be exactly conserved during time evolution, or that the divergence of the magnetic field remains exactly zero. ``Exactly'' here means up to error introduced by floating point precision.

\subsection{Parallelism}

\carpetx provides three levels of parallelism, suitable for modern systems ranging from laptops to high-end supercomputers: shared memory parallelism (multi-threading), accelerators (aka GPUs), and distributed memory parallelism. These mechanisms are implemented in \amrex.

\subsubsection{Shared Memory Parallelism (Multi-Threading)}

\carpetx uses OpenMP~\cite{openmp} for multi-threading. In this setup, a computational grid is allocated as a single entity, and is split into several logical \emph{tiles} for processing. Each tile is handled by one thread. The tile size can be chosen at runtime. We find that efficient tile sizes are very large in the $x$ direction (because each cache line contains multiple grid points that are neighbours in the $x$ direction) and rather small in the $y$ and $z$ directions (since our evolution system contains many variables that quickly fill the cache). We find that a typical efficient tile size would be e.g. $1000\times4\times4$ grid points, where the value $1000$ means that the tile is in practice as large as the grid block in the $x$ direction. 

The routines acting on tiles are scheduled as \codename{OpenMP} \emph{tasks} and then execute independently. When necessary e.g. for \emph{synchronization} (ghost zone exchange) or I/O, a barrier is introduced to ensure all scheduled tasks have finished. 

\subsubsection{Accelerators (GPUs)}
Most new high-end supercomputers require codes to make efficient use of GPUs or other accelerators. Accelerators on such systems provide the majority of the computing power. It is clearly important to be able to use GPUs efficiently on such systems. 

One commonly used approach to do so is to ensure that all data are stored on the accelerator's built-in memory at all times, and are moved from and to the CPU memory only for I/O. It is thus, unfortunately, necessary that all routines of a simulation code run on the accelerator. Copying data to the CPU memory for even a simple task (e.g. to find the maximum of the density) is prohibitively slow; it is more efficient to run such a task on an accelerator. 

The goal of \carpetx is thus to make it easy to write code that runs on accelerators, even if the code would not run efficiently there. Fortunately, code that has been written to run on accelerators will usually also run efficiently on CPUs. 

The loop kernels running on an accelerator are scheduled and then execute as independent tasks. A barrier at the end ensures the accelerator has finished processing tasks before synchronization or I/O, and to ensure tasks execute in a correct order.

\subsubsection{Distributed Memory Parallelism (Message-Passing)}

For distributed memory parallelism, i.e. to run across several compute nodes simultaneously, \amrex offers parallelization via MPI. The rectangular grids which make up the coarse grid and all refined grids are called \emph{blocks} in \amrex. Each grid block is surrounded by a layer of ghost zones that are filled by copying from other MPI processes that may run on different nodes. 

Overall, ghost zones are either filled via communication from another process or via prolongation from a coarser grid. This synchronization needs to be explicitly scheduled by the application code. Grids are automatically restricted (see above) at the same time they are synchronized. 

In our implementation of a flux-conservative hydrodynamics scheme, the state vector consists of conserved quantities that are stored at cell centers. From these, we calculate the fluxes between the cells. These fluxes live on the faces between the cells. The fluxes are synchronized in a conservative manner, ensuring that the fluxes are consistent between all grid blocks, and also between coarse and fine grids. The fine grid fluxes are restricted to coarser grids. This restriction averages the fluxes, so that the integrated fluxes are the same on all refinement levels. 

As we are using a global time step, so-called \emph{refluxing} is not necessary. (Refluxing would keep fluxes between time steps with different step sizes consistent by integrating fluxes in time.)

From the difference of the fluxes, the new state vector is calculated, which is then is also synchronized.

\section{Tests}
\label{sec:tests}

To verify the correctness of the code we perform a number of tests to ensure that the numerical methods implemented are correct and robust. We perform a number of standard tests with a variety of initial conditions and compare the results with analytical solutions as well as results from other codes reported in literature. In this section we describe various tests in gradually increasing level of complexity, ranging from one-dimensional shocktube tests in static Minkowski spacetimes to a three-dimensional magnetized TOV star in dynamical spacetime. 

\subsection{1D MHD shocktube}\label{balsara}
Planar \MHD shocktube tests serve as the primary tests for magnetohydrodynamic codes because they are simple to implement yet provide an excellent testbed for the ability of the code to capture shocks and \MHD-wave structures. We perform five different shocktube tests using the initial conditions proposed by Balsara~\cite{Balsara_2001}, namely "Balsara 1", "Balsara 2", "Balsara 3", "Balsara 4" and "Balsara 5". Balsara 1 is the relativistic generalisation of the shocktube test problem originally proposed by Brio and Wu~\cite{Brio:1988, Putten:1995}. Balsara 2 and 3 are blast wave test problems, with the difference being their initial pressure difference. Balasara 2 has a moderate initial pressure difference ($P_{\rm{left}}/P_{\rm{right}} = 30$)  while Balsara 3 has a very strong initial pressure difference ($P_{\rm{left}}/P_{\rm{right}} = 10^4$). Balsara 4 constitutes a strongly relativistic test problem where two streams with high Lorentz factor ($W \approx 22.37$) collide with each other.  

We perform the tests along x, y and z directions independently, however we show the results only for the x-direction. y- and z-directions give the same result. For the x-direction, we divide the domain in ``left'' and ``right'' states, with ``left'' state being the region $x < 0$ and ``right'' state being the region $x > 0$. This means $x = 0$ serves as the plane of discontinuity for the Riemann problem. For each test, we have used the $\Gamma$-law (ideal gas) EOS given by $P = (\Gamma-1)\rho\epsilon$, with $\Gamma = 2$ for Balsara 1 and $\Gamma=5/3$ for others. We divide the domain between $-0.5$ and $0.5$ in 1600 points which gives a resolution of $\Delta x = 1/1600$. We perform all the tests without AMR and without constrained transport (see~\cite{Moesta_GRHydro_2013} for a detailed explanation of why this setup maintains the divergence-free constraint). We use fourth-order Runge-Kutta (RK4) for time integration with a high Courant-Friedrichs-Lewy (CFL) factor of 0.8 (i.e. $\Delta t/\Delta x = 0.8$). We use von Neumann boundary conditions i.e. we set the flux at the boundary points to be zero and "copy" the data from the nearest point in the interior to the points at the boundary. We perform reconstruction of primitive variables using the TVD reconstruction with the minmod limiter and use the HLLE Riemann solver. We have used a static Minkowski spacetime. We set the left and right states using the values tabulated in Table~\ref{tab:balsara} and evolve the system until $t = t_{\rm{ref}} = 0.55$ for Balsara 5 and $t = t_{\rm{ref}} = 0.4$ for others. We then compare the results obtained at time $t_{\rm{ref}}$ with the exact solution. We have obtained the exact solution from~\cite{giacomazzo_rezzolla_2006}. This test took 29 seconds to evolve the $1600 \times 8 \times 8$ grid for 800 iterations on an NVIDIA RTX A6000 GPU.    

\begin{table}
	\centering
	\caption{ Parameters used for the 1D \MHD shocktube tests. We perform the tests in $x$-direction such that ``Left'' state means the state at $x<0$ and ``Right'' state means the state at $x>0$. $\Gamma$ is the adiabatic index for the ideal gas equation of state, $t_{\rm{ref}}$ is the time at which we report the results, $\rho$ is the density, $\epsilon$ is the specific internal energy, $\vec{v} = (v^x, v^y, v^z)$ is the velocity of the fluid and $(B^x, B^y, B^z)$ are the $x$, $y$ and $z$ components of the magnetic field respectively. }
	\label{tab:balsara}
	\begin{tabular}{|c|c|c|c|c|c|c|c|c|} 
	    \hline
		Test name  & $B^x$ & $\Gamma$ & $t_{ref}$ & State & $\rho$ & $\epsilon$ & $\vec{v}$ & $(B^y, B^z)$  \\
		\hline
		Balsara 1 & 0.5 & 2 & 0.4 & Left & 1.0 & 1.0 & $\vec{0}$ & (1.0, 0)  \\
		 & &  &  & Right & 0.125 & 0.8 & $\vec{0}$ & (-1.0, 0)  \\
		\hline
	    Balsara 2 & 5.0 & 5/3 & 0.4 & Left & 1.0 & 45.0 & $\vec{0}$ & (6.0, 6.0)  \\
		 & &  &  & Right & 1.0 & 1.5 & $\vec{0}$ & (0.7, 0.7)  \\
		\hline
		Balsara 3 & 10.0 & 5/3 & 0.4 & Left & 1.0 & 1500.0 & $\vec{0}$ & (7.0, 7.0)  \\
		 & &  &  & Right & 1.0 & 0.15 & $\vec{0}$ & (0.7, 0.7)  \\
		\hline
		Balsara 4 & 10.0 & 5/3 & 0.4 & Left & 1.0 & 0.15 & (0.999, 0, 0) & (7.0, 7.0)  \\
		 & &  &  & Right & 1.0 & 0.15 & (-0.999, 0, 0) & (-7.0, -7.0)  \\
		\hline
		Balsara 5 & 2.0 & 5/3 & 0.55 & Left & 1.08 & 1.3194 & (0.4, 0.3, 0.2) & (0.3, 0.3)  \\
		 & &  &  & Right & 1.0 & 1.5 & (-0.45, -0.2, 0.2) & (-0.7, 0.5)  \\
		 \hline
	\end{tabular}
\end{table}

\begin{figure}
    \includegraphics[width=0.45\columnwidth]{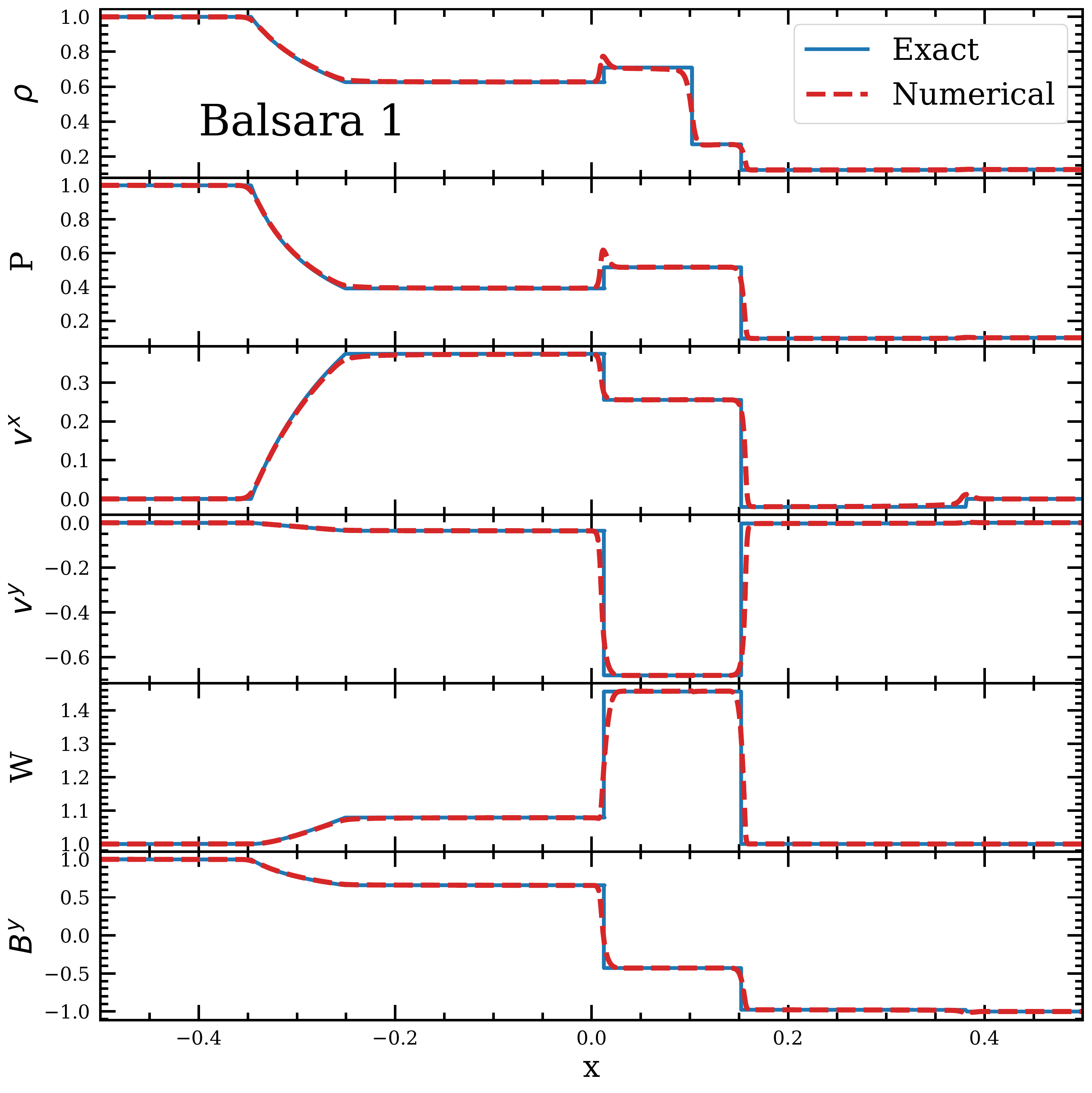}
 	\includegraphics[width=0.45\columnwidth]{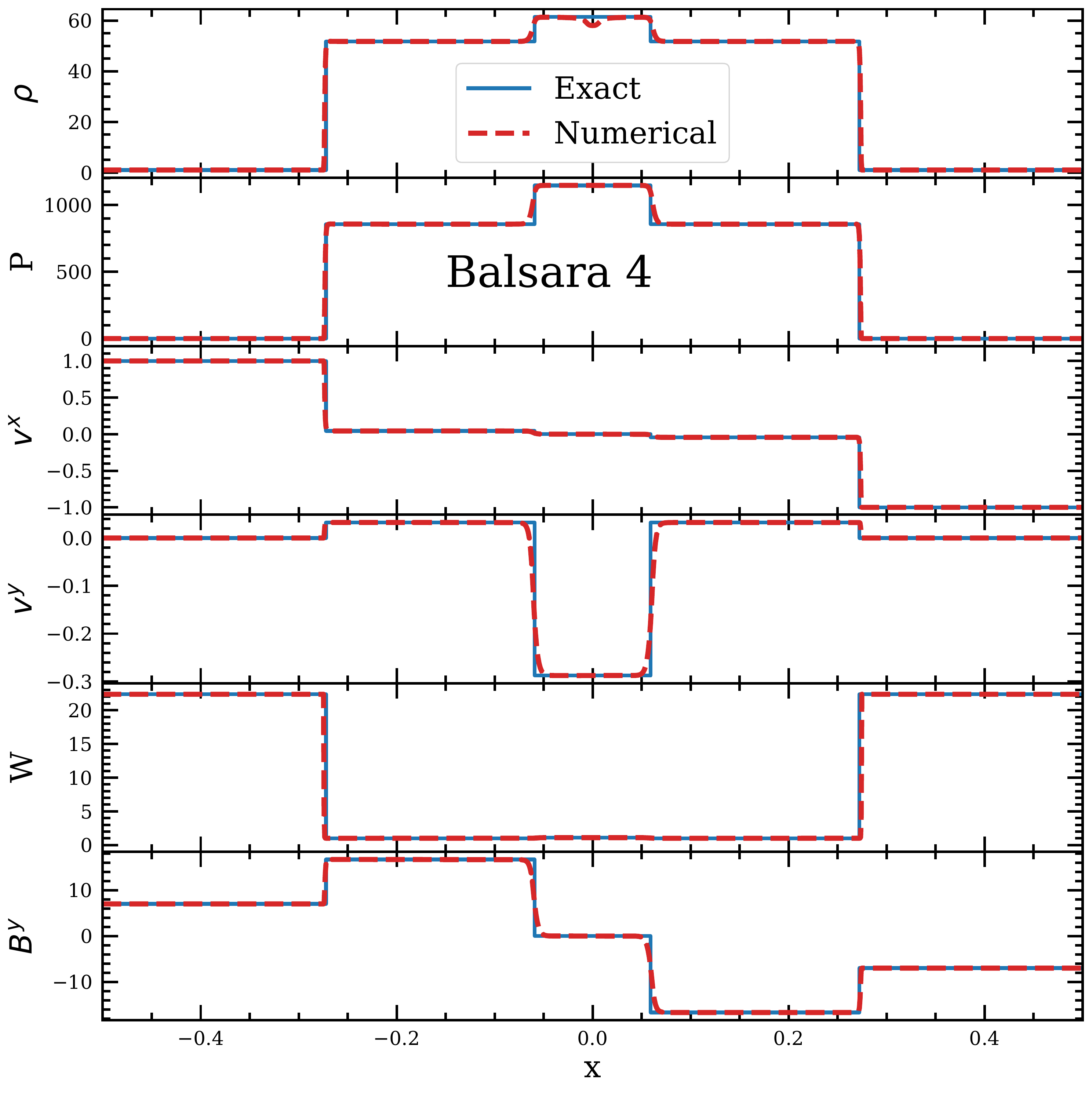}
    \caption{Results for the evolution of 1D \MHD shocktube tests Balsara 1 and Balsara 4 in the left and the right panel respectively. We plot the density $\rho$, pressure $P$, normal velocity $v^x$, tangential velocity $v^y$, Lorentz factor $W$ and tangential component of the magnetic field $B^y$ at time $t_{\rm{ref}}=0.4$ for both tests. We choose the initial conditions as listed in Table~\ref{tab:balsara} on a single-level grid between $[-0.5,0.5]$ with a spatial resolution $\Delta x = 1/1600$ and evolve with a time resolution $\Delta t = 5 \times 10^{-4}$. We find that our results agree well with the analytical solution~\cite{giacomazzo_rezzolla_2006} (limited by numerical errors) and any minor deviations are consistent with those reported by other codes~\cite{Balsara_2001, Giacomazzo_2007, Moesta_GRHydro_2013}. This test took 29 seconds to evolve on an NVIDIA RTX A6000 GPU.} 
    \label{fig:balsara14}
\end{figure}

We show the results for Balsara 1 and Balsara 4 in the left and the right panel of Fig.~\ref{fig:balsara14} respectively. We find that the results are in very good agreement with the analytical solution (limited only by numerical errors) and also agree well with other results reported in literature~\cite{Giacomazzo_2007, Zanna:2007, Anderson_2006, Anton_2006, Balsara_2001, giacomazzo_rezzolla_2006, Balsara_1998, Zanna:2003}. In the left panel of Fig.~\ref{fig:balsara14} we plot the density, pressure, $x$-velocity, $y$-velocity, Lorentz factor and $y$-component of the magnetic field for Balsara 1.  We find that a reasonably relativistic flow with a Lorentz factor of $\sim 1.458$ has developed from the initial discontinuity and the code has captured all the elementary waves which include a left-going fast rarefaction wave, a left-going compound wave, a contact discontinuity, a right-going slow shock and a right-going fast rarefaction wave. This is also in good agreement with the exact solution. In the right panel of Fig. \ref{fig:balsara14} we plot the same quantities for the \MHD collision problem Balsara 4. The flow develops two very strong fast shocks and two slow shocks, each going to the right and left respectively. We find deviations from the exact result at the center but they are on the same level as~\cite{Balsara_2001, Giacomazzo_2007, Moesta_GRHydro_2013}. 

\begin{figure}
 	\includegraphics[width=0.30\columnwidth]{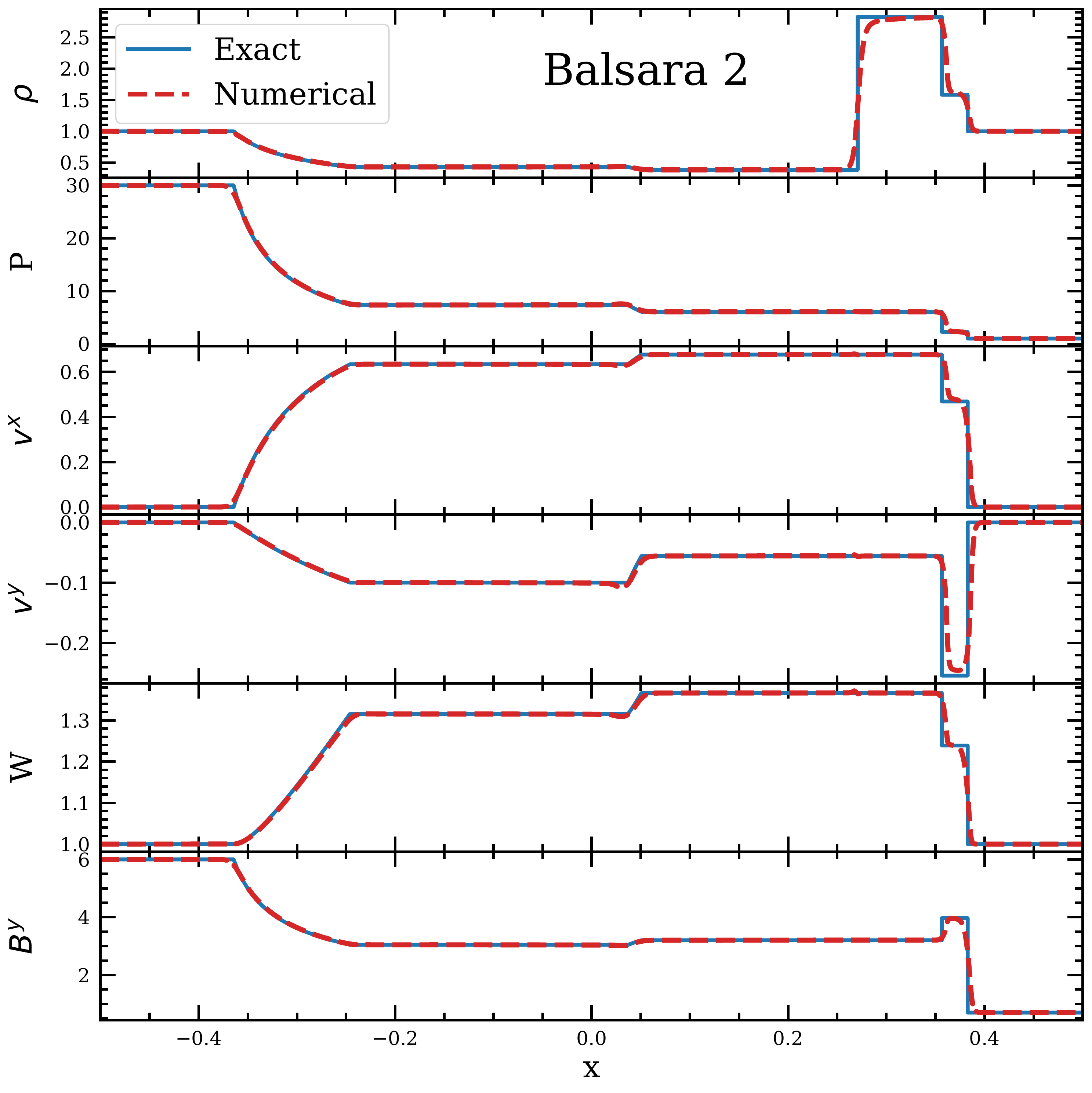}
 	\includegraphics[width=0.30\columnwidth]{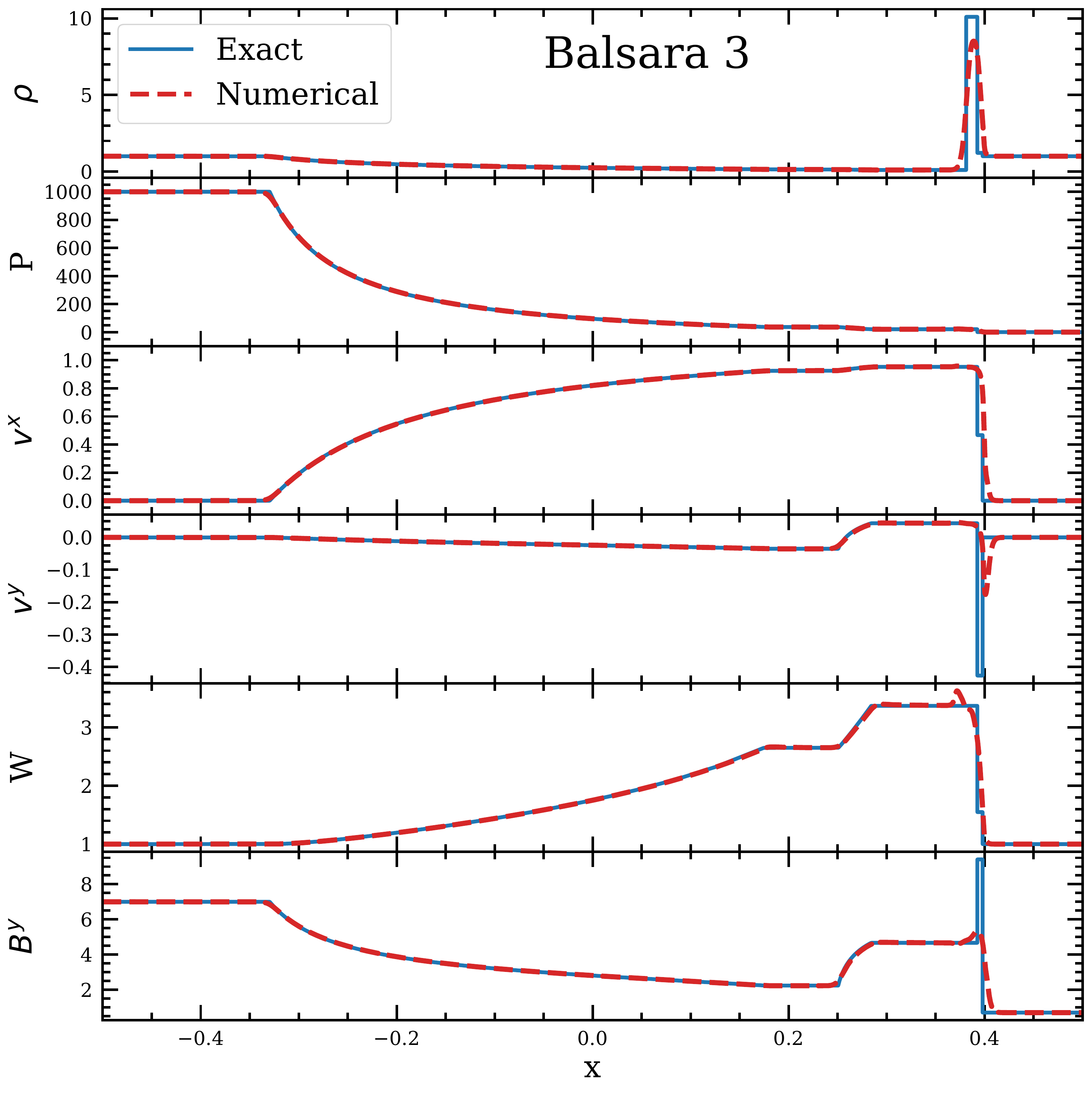}
 	\includegraphics[width=0.30\columnwidth]{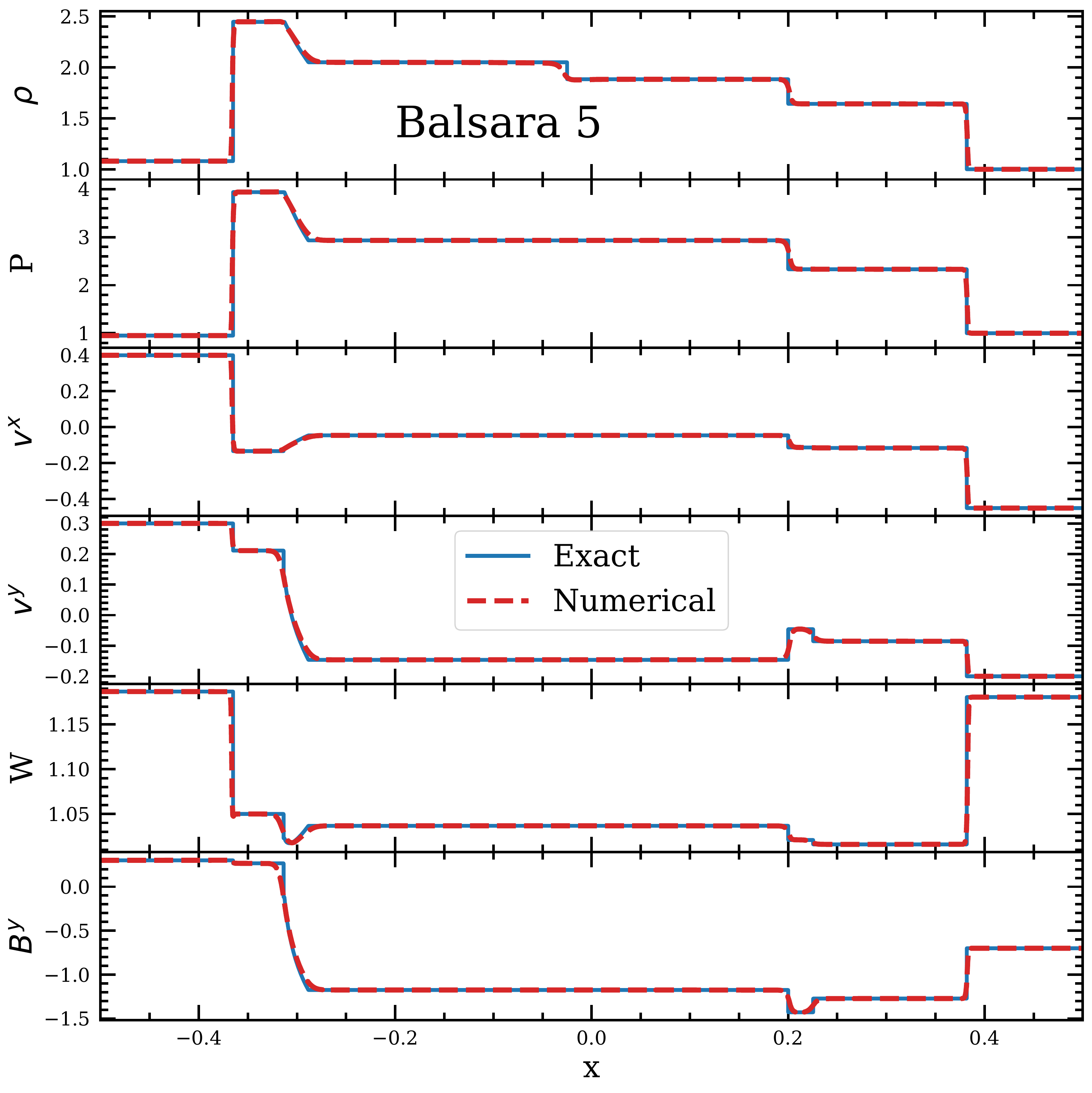}
    \caption{Same as Fig.~\ref{fig:balsara14} but for Balsara 2, Balsara 3 and Balsara 5. We plot the results at time $t_{\rm{ref}}=0.4$ for Balsara 2 and 3, and at time $t_{\rm{ref}}=0.55$ for Balsara 5. We find excellent agreement with the analytical results~\cite{giacomazzo_rezzolla_2006} for Balsara 2 and Balsara 5. Balsara 3 is the most demanding of all Balsara tests and we find that there is a  considerable undershoot near the right moving fast shock for $\rho$, $v^y$ and $B^y$. However, using WENO reconstruction instead of TVD reconstruction at the same resolution resolves this issue to a large extent and makes the deviations comparable to those observed by~\cite{Balsara_2001, Moesta_GRHydro_2013}}.
    \label{fig:balsara345}
\end{figure}

We also show the results for Balsara 2, Balsara 3 and Balsara 5 in Fig~\ref{fig:balsara345}. For more details about the physics that these tests are testing please see~\cite{Moesta_GRHydro_2013}. For Balsara 2 and Balsara 5, we reproduce the analytic results very well. Balsara 3 is the most challenging of all the Balsara tests and we find that the result shows a considerable undershoot near the right moving shock at the current resolution of $\Delta x = 1/1600$. This however gets resolved to a large extent when we use WENO instead of TVD reconstruction. That makes the deviations comparable to those reported by~\cite{Balsara_2001, Moesta_GRHydro_2013}. We report the results only for TVD reconstruction for consistency among all the tests.

\subsection{2D cylindrical explosion}
We next perform a two dimensional test which involves a strong cylindrical explosion to test the ability of our code to capture an expanding blast wave in multiple dimensions. This test is known to push the limits of an \MHD scheme and spot well-hidden bugs because of the simplicity of the setup. Unlike the Balsara tests we do not have an exact solution for this test, hence we compare our results with results of other codes available in literature. We use the test setup described in~\cite{Moesta_GRHydro_2013} which is originally based on test setup of~\cite{Komissarov:1999} for the relatively weak magnetic field case ($B^x=0.1$).  

We perform the test in the $x$-$y$ plane with the density profile given by
\begin{eqnarray}
\rho(r) = \left\{\begin{array}{lll}
\rho_{\rm in} & ; & r\le r_{\rm in}\,\,,\\
\exp\left[\frac{(r_{\rm out}-R)\ln\rho_{\rm out}+(r-r_{\rm in})\ln\rho_{\rm in}}{r_{\rm out}-r_{\rm in}}\right] & ; & r_{\rm in}<r<r_{\rm out}\,\,,\\
\rho_{\rm out} & ; & r\ge r_{\rm out}\,\,,
\end{array}\right.
\end{eqnarray}
where $r = \sqrt{x^2+y^2}$ is the radial distance from the origin and ($r_{\rm in}$, $r_{\rm out}$) are radial parameters. The pressure gradient has an equivalent form where $\rho_{\rm in}$ and $\rho_{\rm out}$ are replaced by $P_{\rm in}$ and $P_{\rm out}$ respectively. We set the initial fluid velocity to zero in all directions and use a uniform magnetic field along $x$-direction. The values of the parameters are as follows:
\begin{equation}
r_{\rm in}=0.8,~r_{\rm out}=1.0;~\rho_{\rm in}=10^{-2},~\rho_{\rm out}=10^{-4};~P_{\rm in}=1.0,~P_{\rm out}=3 \times 10^{-5};~B^i = (0.1,0,0).
\end{equation}
Thus there is a large pressure jump of $\sim10^{5}$ going from $r_{\rm out}$ to $r_{\rm in}$. We use a $200\times200\times8$ grid with $x$- and $y$-directions spanning the coordinate range $[-6,6]$. This results in a resolution of $\Delta x = \Delta y = 0.06$. We use the $\Gamma$-law EOS with $\Gamma = 4/3$ and use constrained transport for the magnetic field evolution to conserve the divergence-free constraint of the magnetic field. We use von Neumann boundary conditions as described in section~\ref{balsara} and do not use mesh refinement. We use TVD reconstruction with the minmod limiter, the HLLE Riemann solver and RK4 for time integration with the CFL factor of 0.25. With this numerical setup, we run the simulation until $t=4$ and compare the two-dimensional and one-dimensional profiles of different physical quantities with previous results. This test took 15 seconds to evolve the $200 \times 200 \times 8$ grid for 267 iterations on an NVIDIA RTX A6000 GPU.  

\begin{figure}
    \includegraphics[width=0.45\columnwidth, viewport=30 30 1250 1250]{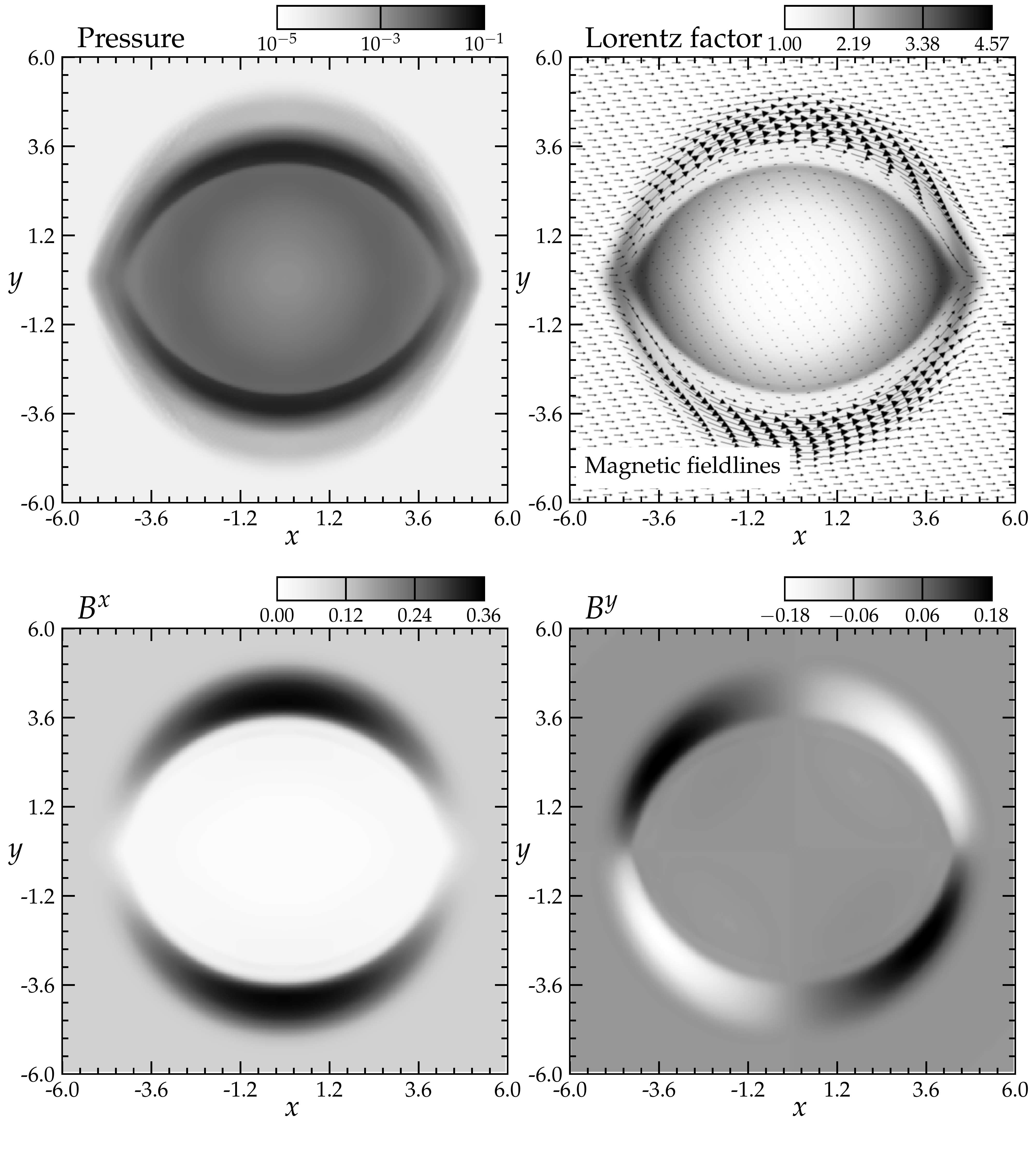}
    \includegraphics[width=0.45\columnwidth]{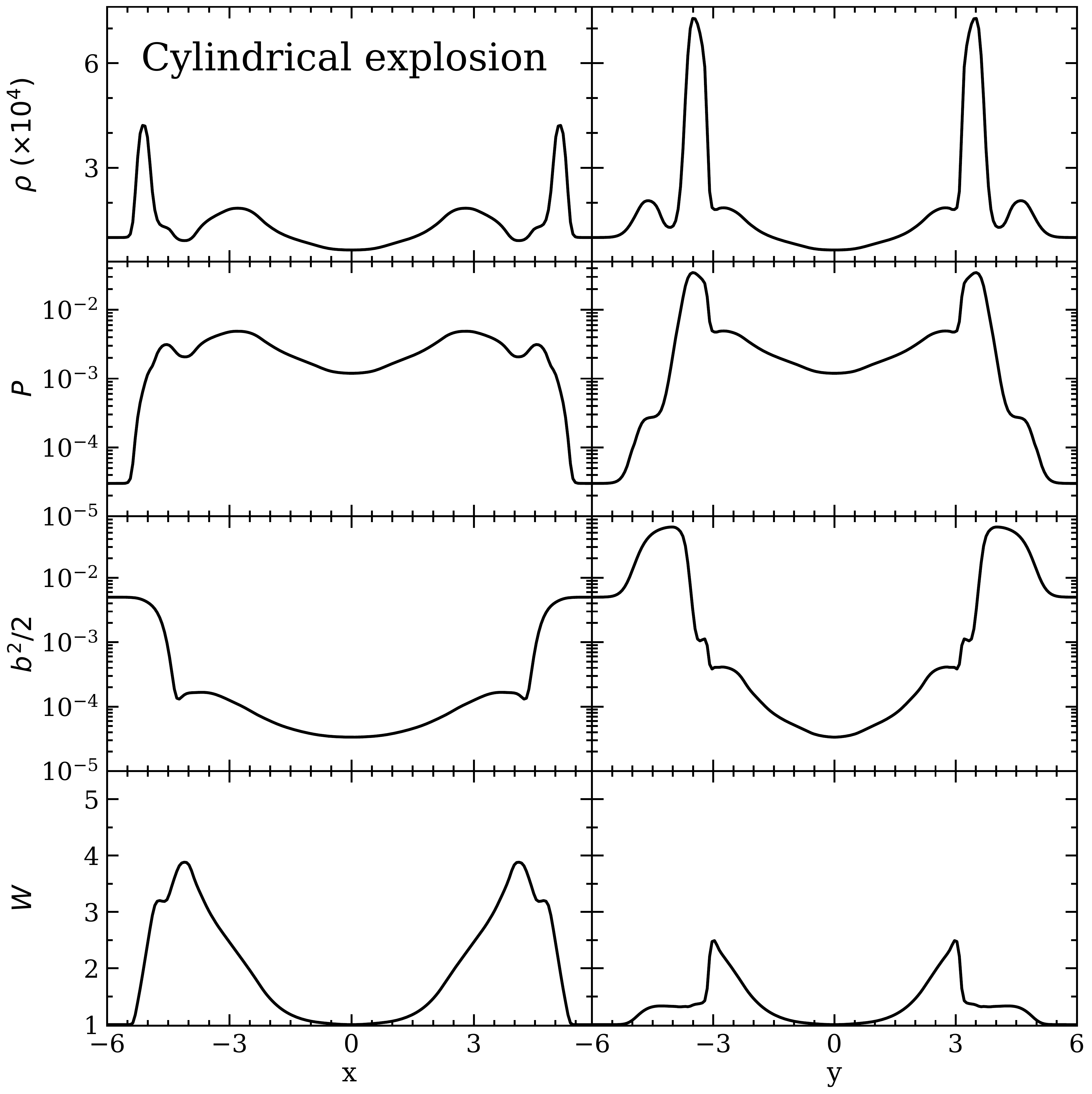}
    \caption{ \textit{Left panel}: The evolved state of the 2D cylindrical explosion test at time $t = 4$. We plot the logarithm of pressure, the Lorentz factor as well as the $x$ and $y$ components of the magnetic field. We also overlay the magnetic field lines on top of the Lorentz factor plot.  We perform the test on a single level grid between $-6 < x, y < 6$ with a spatial resolution of $\Delta x = \Delta y = 0.06$ and evolve with a temporal resolution of $\Delta t = 0.015$. We find that our results agree well with the results reported by~\cite{Moesta_GRHydro_2013} and~\cite{Komissarov:1999} for the same test setup. \textit{Right panel}: 1D profile along $x=0$ and $y=0$, for density $\rho$, pressure $P$, magnetic pressure $b^2/2$ and Lorentz factor $W$. We find overall good agreement but minor deviations with Fig. 6 of~\cite{Moesta_GRHydro_2013} for density and magnetic pressure. These deviations occur because~\cite{Moesta_GRHydro_2013} uses vertex-centering while \theCode uses cell-centering for hydrodynamic variables. The deviations go away when we perform the test with same coordinate discretization as~\cite{Moesta_GRHydro_2013}. This test took 15 seconds to evolve on an NVIDIA RTX A6000 GPU. }
    \label{fig:cylexp}
\end{figure}

We show the two-dimensional profiles for gas pressure $P$, Lorentz factor $W$ as well as the magnetic field in the $x-y$-plane at time $t=4$ in the left panel of Fig.~\ref{fig:cylexp}. We also show the numerically determined magnetic field lines on top of Lorentz factor plot. The plot for $P$ is in logarithmic scale. We find that our two-dimensional results show very good agreement with Fig. 5 of~\cite{Moesta_GRHydro_2013} and Fig. 10 of~\cite{Komissarov:1999}. Next, we plot the one-dimensional profiles by taking slices along $y=0$ and $x=0$ for rest-mass density, gas pressure, magnetic pressure and Lorentz factor at time $t=4$. Gas pressure and magnetic pressure are shown in logarithmic scale. We show these 1D profiles in the right panel of Fig.~\ref{fig:cylexp}. We find good agreement with Fig. 6 of~\cite{Moesta_GRHydro_2013}. Slight deviations from the 1D profiles of~\cite{Moesta_GRHydro_2013} are present because our code uses cell-centering for hydrodynamic variables and vertex-centering for spacetime variables while~\cite{Moesta_GRHydro_2013} uses the same centering for both hydrodynamic and spacetime variables. This leads to slightly different numerical values (limited to numerical precision) for initial density and pressure profiles on the discrete grid due to their coordinate dependence. This small discrepancy is resolved (not shown here) when we use the same coordinate discretization for hydrodynamic variables in~\cite{Moesta_GRHydro_2013} and our code. 

\subsection{2D magnetic rotor}
Next we perform the magnetic rotor test which has high initial Lorentz factors and a strong initial discontinuity. This test was generalized to the relativistic \MHD case in~\cite{Zanna:2003}. Like the cylindrical explosion test, we do not have an analytical solution for this test and hence we compare our results with the results of other codes. The test consists of a cylindrical column of radius $r_{\rm{in}} = 0.1$ with its axis along $z$-direction rotating with an angular velocity of $\Omega=9.95$. Thus the fluid 3-velocity at the outer edge of the cylinder reaches the value $v_{\rm{max}}=0.995$ which corresponds to a high Lorentz factor of $\sim10.012$. The density inside the cylinder is uniform with the value $\rho_{\rm{in}} = 10$ while the medium surrounding the cylinder has a lower uniform density $\rho_{\rm{out}} = 1$. We do not apply any smoothing to the density profile, thus a strong initial discontinuity is present at the edge of the cylinder. The pressure is the same both inside and outside the cylinder with the value $P_{\rm{in}} = P_{\rm{out}} = 1$. A uniform magnetic field along $x$-direction is present both in the interior and exterior of the cylinder with the value $B^x = 1.0$. 

We use a $400\times400\times8$ grid with $x$ and $y$ directions spanning the coordinate range $[-0.5,0.5]$. This gives us a resolution of $\Delta x= \Delta y = 0.0025$. We perform the test on a flat spacetime and without mesh refinement. We use von Neumann boundary condition as described in section~\ref{balsara} and use constrained transport to evolve the magnetic field. We use a $\Gamma$-law equation of state with an adiabatic index $\Gamma=5/3$. We use TVD reconstruction with the minmod limiter and the HLLE Riemann solver. We perform time integration using RK4 with a CFL factor of 0.25. With this numerical setup we evolve the system until $t=0.4$ and again compare the two-dimensional and one-dimensional profiles of different physical quantities with previous results. This test took 97 seconds to evolve the $400 \times 400 \times 8$ grid for 640 iterations on an NVIDIA RTX A6000 GPU.

\begin{figure}
    \includegraphics[width=0.45\columnwidth, viewport=30 30 1250 1250]{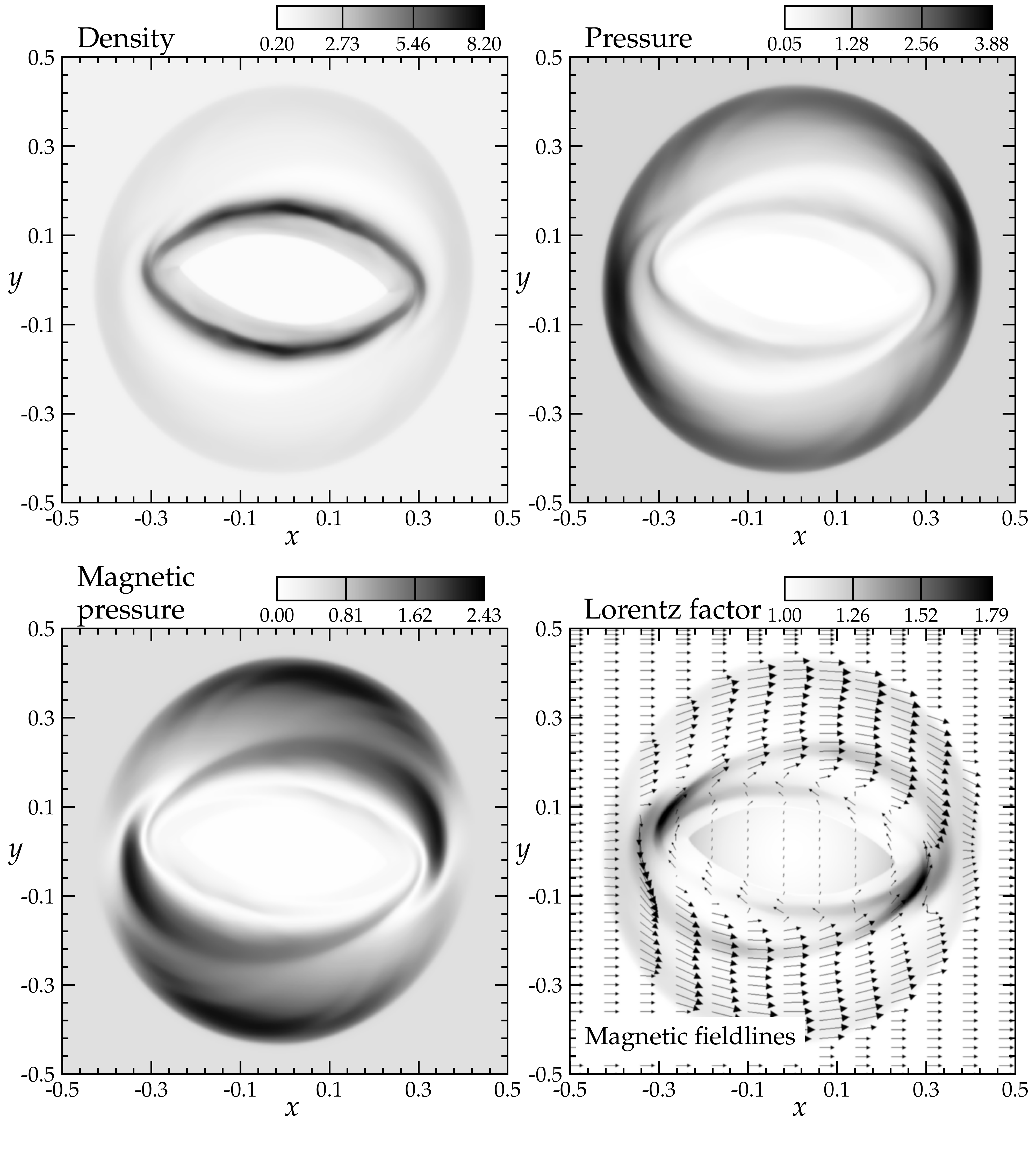}
 	\includegraphics[width=0.45\columnwidth]{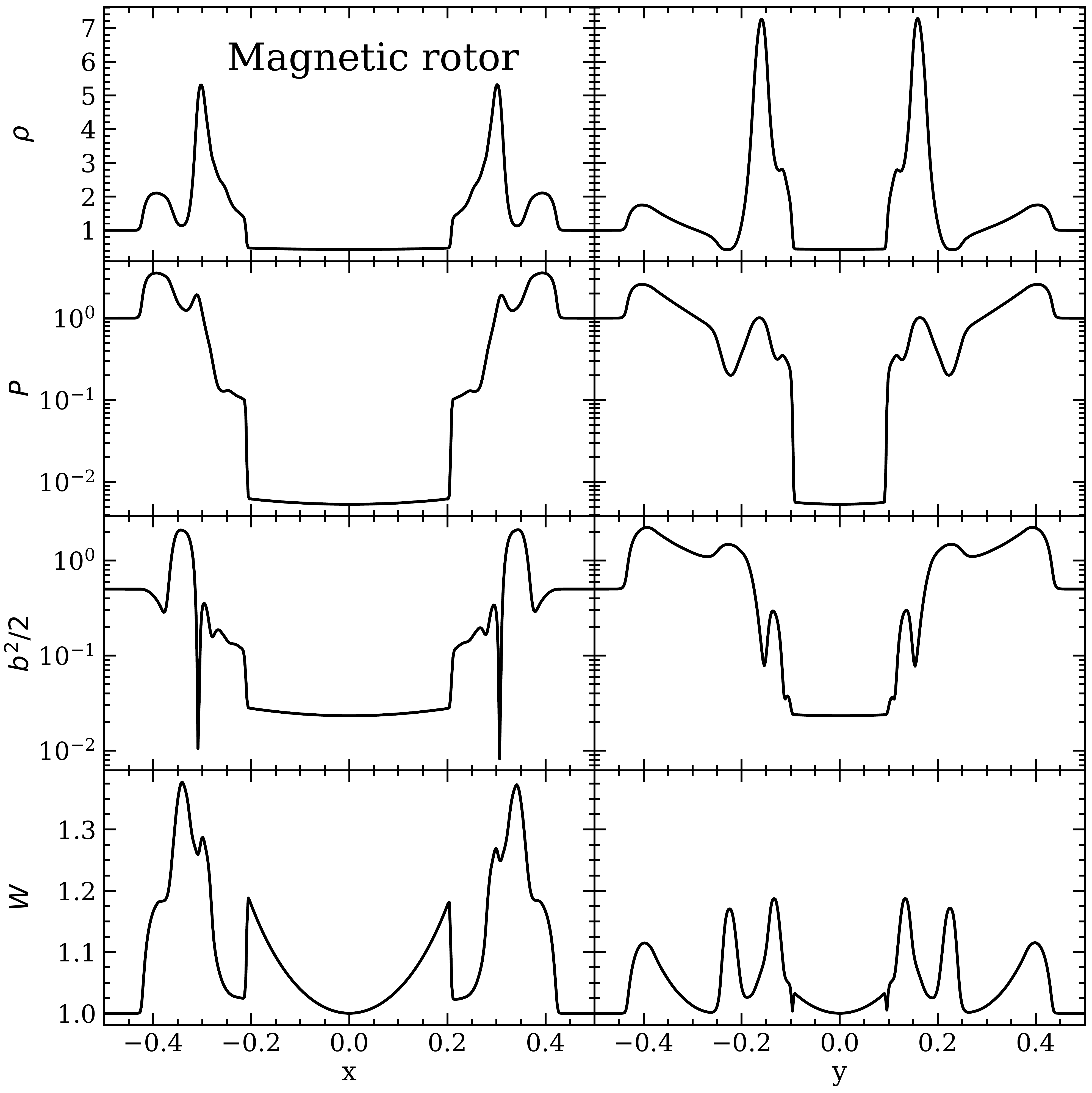}
    \caption{\textit{Left panel}: The evolved state of the 2D magnetic rotor test at time $t = 0.4$. We plot the density, pressure, magnetic pressure and Lorentz factor, and also overlay the magnetic field lines on top of the Lorentz factor plot.  We perform the test on a single-level grid between $-0.5 < x, y < 0.5$ with a spatial resolution of $\Delta x = \Delta y = 1/400$ and evolve with a temporal resolution of $\Delta t = 6.25 \times 10^{-4}$. These 2D results show very good agreement with~\cite{Zanna:2003}, \cite{Etienne:2010} and~\cite{Moesta_GRHydro_2013}. \textit{Right panel}: 1D profiles along $x=0$ and $y=0$ for density $\rho$, pressure $P$, magnetic pressure $b^2/2$ and Lorentz factor $W$. Again, we find excellent agreement overall but some minor deviations with Fig. 6 of~\cite{Moesta_GRHydro_2013} and Fig. 9 of~\cite{Etienne:2010} due to different centering of variables. In fact, these deviations becomes irrelevant when we compare the results of other codes at different resolutions (see Fig. 9 of~\cite{Etienne:2010} for example). This test took 97 seconds to evolve on an NVIDIA RTX A6000 GPU. } 
    \label{fig:rotor}
\end{figure}

In the left panel of Fig.~\ref{fig:rotor} we show the 2D profiles of density, fluid pressure, magnetic pressure and Lorentz factor at $t=0.4$. Magnetic field lines are shown on top of the Lorentz factor plot. As expected, we find that magnetic braking slows down the fluid and this causes the maximum Lorentz factor to drop to $\sim 1.85$ from the initial value of $\sim10$. Magnetic field lines are also dragged by the rotation of the fluid which causes the field lines to rotate by almost 90 degrees in the central region while the field lines remain unchanged at large radii. Material has been swept away from the central region causing the central density to decrease from the initial maximum value of 10 to the minimum value of $\sim 0.4$. Our results show very good agreement with Fig. 5 of~\cite{Zanna:2003}, Fig. 7 of~\cite{Moesta_GRHydro_2013} and Fig. 8 of~\cite{Etienne:2010}. In the right panel of Fig.~\ref{fig:rotor} we show the 1D profiles taken along $y=0$ and $x=0$ for density, fluid pressure, magnetic pressure and Lorentz factor at $t=0.4$. Our results show excellent agreement with Fig. 8 of~\cite{Moesta_GRHydro_2013} and Fig. 9 of~\cite{Etienne:2010}. Again, we find minor deviations from the results of~\cite{Moesta_GRHydro_2013} but these again are due to the difference in centering of variables between~\cite{Moesta_GRHydro_2013} and our code. In fact, these deviations become irrelevant when we take into account the differences in results from the same code at different resolutions, for example Fig. 9 of~\cite{Etienne:2010}.  

\subsection{3D TOV star}
As the final test we simulate a stationary and spherically symmetric neutron star with poloidal magnetic field in three dimensions. This test is more challenging than previous tests because (1) we perform it in three dimensions and (2) this test probes both GR and \MHD. This is our first dynamical spacetime test and we use the Z4c formalism to evolve the spacetime variables. We obtain the neutron star model via the solution of Tolman-Oppenheimer-Volkhoff (TOV) equation~\cite{Tolman:1939} and set up a poloidal magnetic field on top of this fluid configuration. The solution is stationary, hence this test helps us to investigate the ability of our code to maintain this stationary configuration during evolution in a dynamical spacetime setup. 

The test involves setting up self-consistent initial data for fluid and spacetime variables, and then evolving the hydrodynamic variables with \theCode and the spacetime variables using the Z4c formalism with \texttt{Z4c}.  In order to calculate the initial data, we solve the TOV equation in 1D using a polytropic equation of state with polytropic constant $K = 100$, adiabatic index $\Gamma=2$ and initial central density of $1.28\times 10^{-3} M_{\odot}$. These parameters form a TOV star with mass $M = 1.4 M_\odot$ and radius $R = 8.125 M_\odot $. We obtain the initial data for poloidal magnetic field using the vector potential $A_\varphi =
A_b\,\varpi^2 (1 - {\rho_0}/{\rho_0^{\rm max}})^{n_p}
\max{(P-P_{\rm cut},0)}$ with $\varpi \equiv \sqrt{(x-x_\star)^2 +
  y^2}$, where $A_b$ determines the strength of the
initial magnetic field, $n_p$ shifts the location of maximum initial magnetic field, $P_{\rm cut}$ is the cut-off pressure below which we set magnetic field to zero and $(x_\star, 0)$ is the location of center of the star. For the test we choose $A_b = 1.0$, $n_p = 0$ and $P_{\rm cut} = 10^{-6}$. We use an artificial atmosphere outside the star where we set the density to $10^{-10} M_{\odot}$ and velocity to zero. We then calculate the pressure and specific energy density in the atmospheric region using a polytropic equation of state with $K \simeq 100$ and $\gamma = 2$. We interpolate the one-dimensional TOV initial data to the three dimensional grid and use it as initial data for the test.    

We set up four simulations on multiple AMR levels with fine grid resolutions of $1M_\odot$($r0$), $0.5M_\odot$($r1$), $0.25M_\odot$($r2$) and $0.125M_\odot$($r3$). The corresponding coarse grid resolutions are $8M_\odot$($r0$), $8M_\odot$($r1$), $8M_\odot$($r2$) and $4M_\odot$($r3$) respectively. The extent of the outermost cube is $640M_\odot$ for all simulations while the extent of the innermost cube is $96M_\odot$($r0$), $48M_\odot$($r1$), $24M_\odot$($r2$) and $22M_\odot$($r3$) respectively. We perform the simulations in the entire domain without the use of any symmetry. We evolve the system with a $\Gamma$-law EOS and $\Gamma=2.0$, the HLLE Riemann solver, 5th order WENO reconstruction and an RK4 time integrator. We use the CFL factor ($\Delta t/ \Delta x$) of 0.25 where $\Delta x$ is the resolution of the finest grid. We use constrained transport for the magnetic field. For the spacetime evolution using \texttt{Z4c}, we use the values for constraint damping parameters $\kappa_1 = 0.02$ and $\kappa_2 = 0.0$, and dissipation coefficient of $0.32$~\cite{Bernd:2008}. The finite numerical resolution acts as a perturbation to the equilibrium configuration of the star which leads to oscillations. The frequencies of these oscillations can be measured by taking a Fourier transform of the time variation of the central density of the star and then compared with the frequencies calculated using linear perturbation theory~\cite{Yoshida:2001, Font:2002}. These frequencies are different for static vs dynamical spacetime evolution and modes of gravity can only be obtained when the test is run with dynamical spacetime.

\begin{figure}
 	\includegraphics[width=0.45\columnwidth]{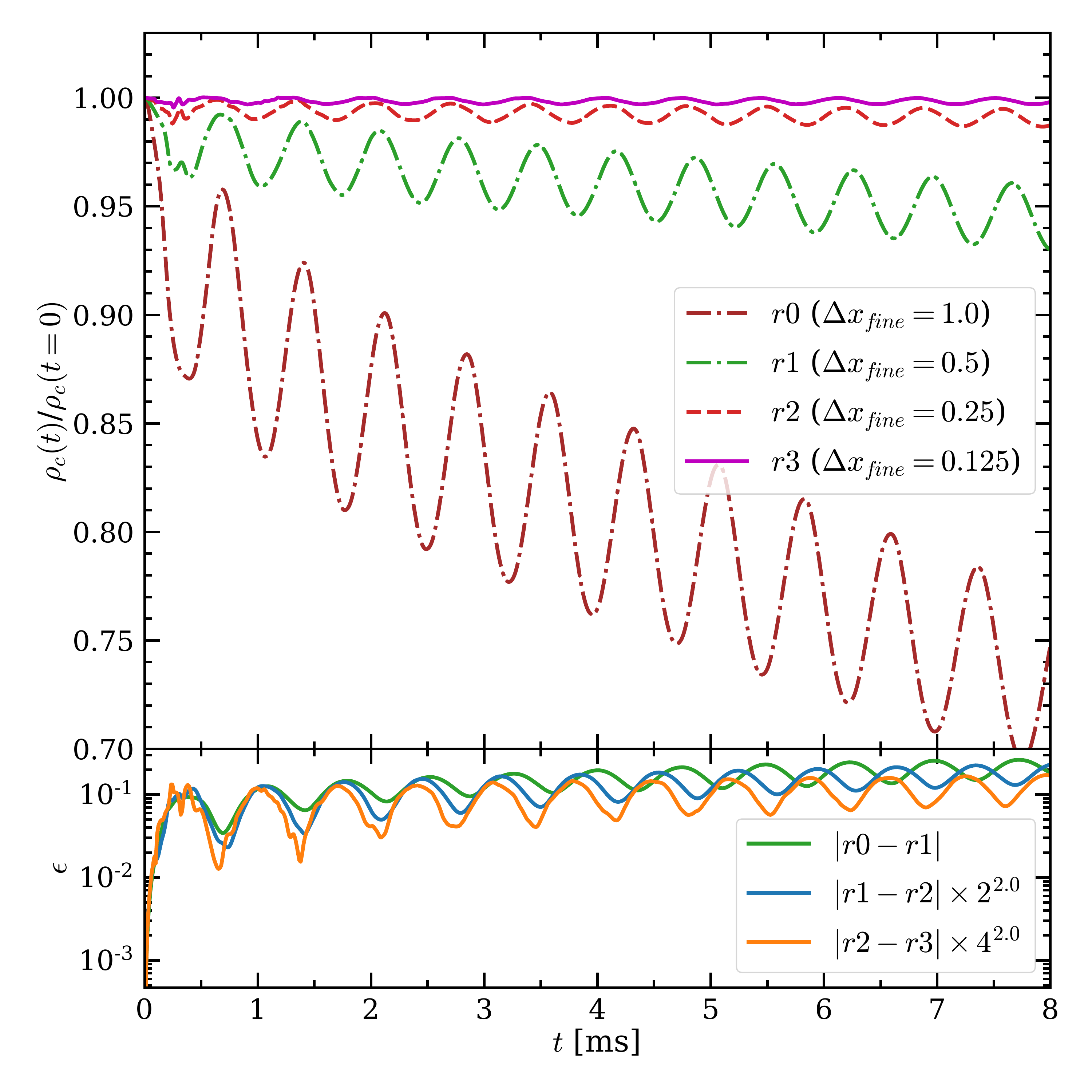}
 	\raisebox{-0.04\height}{\includegraphics[width=0.45\columnwidth]{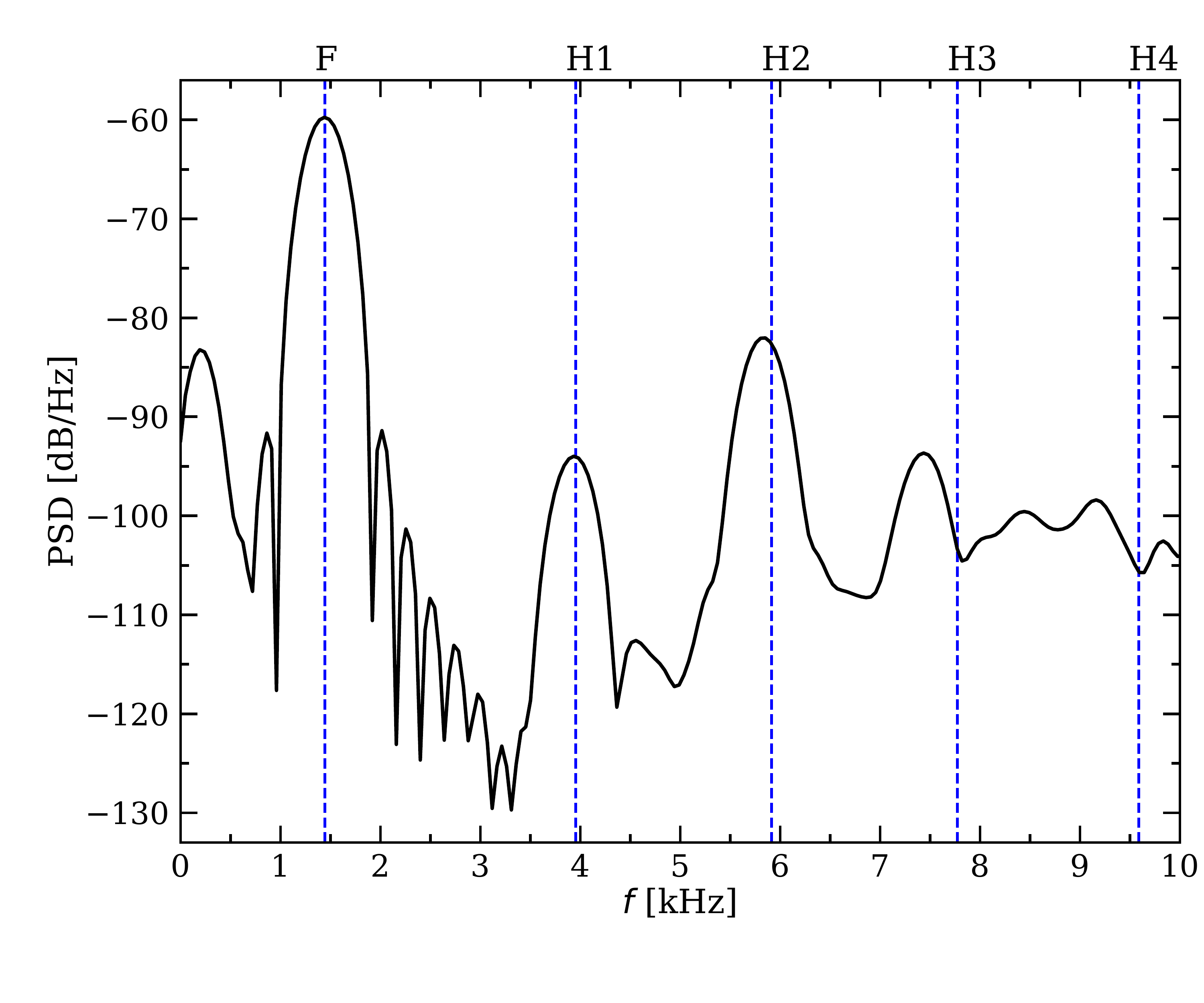} }
    \caption{\textit{Left Panel:} Normalized central density as a function of time for resolutions $r0 (1M_\odot)$, $r1 (0.5M_\odot)$, $r2 (0.25M_\odot)$ and $r3 (0.125M_\odot)$ until $t = 8ms$, where $\Delta x_{\rm{fine}}$ represents the resolution of the finest grid. Boundaries are present at $\pm 320M_\odot$. We find that the solution converges to the analytic solution (constant density) as we increase the resolution. To quantify this we show the time variation of absolute difference $\epsilon$ between $\{r0, r1\}$, $\{r1, r2\}$ and $\{r2, r3\}$ at the bottom. This gives us a convergence order of $\sim 2$. \textit{Right Panel:} Power spectral density (PSD) of the oscillations of the central density of the star. We also show the frequencies of oscillations calculated using perturbation theory for comparison.  We find good agreement with the fundamental frequency ($F$) and the first two harmonics ($H1, H2$). The third harmonic ($H3$) appears slightly shifted from the correct value at this resolution. Correct $H3$ and even higher modes can be obtained by increasing the resolution further.} 
    \label{fig:tov}
\end{figure}

We show the results for time variation of the normalized central density $\rho_c$ in the left panel of Fig. \ref{fig:tov}. We find that the central density gets progressively closer to the ideal solution going from resolution $r0$ to $r3$. We also show this using the plot of absolute difference between $\{r0, r1\}$, $\{r1, r2\}$ and $\{r2, r3\}$ as a function of time. We get a convergence order of $2$ going from low to high resolution. The amplitude of oscillations is larger in this case when compared to the results of \grhydro with the BSSN formalism. This happens because the centering is different for hydrodynamic and spacetime variables in \theCode. The stress-energy tensor $T_{\mu\nu}$ acts as the coupling term between hydrodynamic and spacetime variables. $T_{\mu\nu}$ itself is vertex-centered just as spacetime variables, but hydrodynamic variables are cell-centered. $T_{\mu\nu}$ depends on both hydrodynamic and spacetime variables and hence we have to interpolate the values of hydrodynamic variables from cell centers to vertex centers, which means that $T_{\mu\nu}$ is only an approximate value in \theCode and not an exact value as in~\cite{Moesta_GRHydro_2013}. This leads to higher numerical errors thus causing larger amplitude of oscillations. We have used 4th order 3D interpolation of fluid variables in the calculation of $T_{\mu\nu}$  because we found that 2nd order interpolation (i.e averaging) leads to an even larger, unacceptable amplitudes of oscillations. In the right panel of Fig.~\ref{fig:tov} we show the power spectral density (PSD) of the oscillations of the central density for our highest resolution simulation ($r3$). We obtain the PSD in a manner similar to~\cite{Loffler_2012}. We also plot the frequencies obtained using perturbation theory against the PSD. We find clear agreement between the fundamental mode ($F$) as well the first ($H1$) and second harmonic ($H2$). The third harmonic ($H3$) appears to be shifted slightly at this resolution. At higher resolutions we should obtain the higher modes too, although at a substantially higher computational cost. 

\section{Performance and Scalability}
\label{sec:perf}
We test the performance of \theCode mainly on OLCF's supercomputer Summit. Each compute node on Summit consists of 6 NVIDIA Tesla V100 GPUs with 16 GB of High Bandwidth Memory per GPU. The amount of memory per GPU determines the maximum limit on the number of cells per GPU that can be used during a simulation. The computation/communication ratio should be kept high to use the GPUs efficiently and reduce the effect of memory transfers on the performance of the code.  

In all the benchmarks that we report hereafter, we have used a periodic boundary condition, 5th order WENO reconstruction, and 3 ghost zones. We do not write any output files during runtime and do not include the time required to set up the initial condition in the benchmarks. We only report the performance results for unigrid setups (i.e. without AMR) in the current work.

We perform the benchmark for 3 different physics setups:
\begin{enumerate}
    \item \texttt{SetupA}: static spacetime + ideal gas equation of state + 2nd order Runge-Kutta (RK2) time integrator 
    \item \texttt{SetupB}: static spacetime + \textit{tabulated} equation of state + \textit{4th} order Runge-Kutta (RK4) time integrator 
    \item \texttt{SetupC}: \textit{dynamic} spacetime + tabulated equation of state + 4th order Runge-Kutta time integrator
\end{enumerate}
We use \texttt{setupA} mainly to compare our performance with other codes while \texttt{setupB} and \texttt{setupC} are our production simulation setups in a static and a dynamic spacetime, respectively. We report the results in terms of zone-cycles/s which we calculate using ($n_{\rm{iter}} \times n_{\rm{cells}})/t$, and zone-cycles/s/GPU which we calculate using ($n_{\rm{iter}} \times n_{\rm{cells}})/(t \times n_{\rm{GPU}})$,  where $t$ is the total time in seconds taken to finish $n_{\rm{iter}}$ iterations on a grid with $n_{\rm{cells}}$ cells running on $n_{\rm{GPU}}$ GPUs. Thus a single zone-cycle in the  calculation above includes all intermediate steps for RK2/RK4 time integrators. Zone-cycles/s measures how efficient the simulation is performing overall (wall-time efficiency), whereas zone-cycles/s/GPU measures how efficient the simulation is performing on each GPU (cost efficiency).

\subsection{Single GPU Benchmarks}
We first present the results for single GPU benchmarks (i.e. without communication) for the different physics setups. For \texttt{setupA} on a single Summit V100 GPU, we get $0.44\times10^8$ and $0.47\times10^8$ zone-cycles/s running with $152^3$ cells/GPU and $240^3$ cells/GPU respectively. This is comparable with the single summit V100 GPU performance of H-AMR~\cite{HAMR_Liska} which obtains $\sim0.85\times10^8$ zone-cycles/s for a similar physics setup. For \texttt{setupB}, the performance drops to $0.24\times10^8$ and $0.26\times10^8$ zone-cycles/s running with $152^3$ cells/GPU and $240^3$ cells/GPU respectively. This drop of $\sim1.8$ times in speed occurs because we use tabulated EoS and RK4 integrator which has 4 (twice as many) intermediate steps per zone-cycle. Table lookups/interpolations are also more expensive computationally.

For \texttt{setupC}, we get $0.062\times10^8$ zone-cycles/s on a single GPU running with $168^3$ cells/GPU. We cannot use more than $168^3$ cells/GPU in this case because of the increased GPU memory consumption by the additional spacetime variables. The performance drops by a factor of $\sim4$ compared to the static spacetime case due to additional computations for spacetime evolution, but also partly due to the fact that our spacetime solver isn't fully optimized for GPUs yet. (For example, there are a few instances of register spills for which work is being done to resize the kernels.) We expect the single GPU benchmark for this case to be $\sim0.1\times10^8$ zone-cycles/s after optimization. 

We also perform benchmarks on a single NVIDIA RTX A6000 GPU. For \texttt{setupA} we get $0.090 \times 10^8$ zone-cycles/s running on both $152^3$ and $240^3$ cells/GPU which indicates that the compute load gets saturated already at $152^3$ cells/GPU. The performance is $\sim 5$ times slower than that of a V100 GPU. For \texttt{setupB} we obtain $0.045 \times 10^8$ zone-cycles/s at $152^3$ cells/GPU which is $\sim 5.5$ times slower than the performance on V100 GPU. \texttt{setupC} gives us $0.020 \times 10^8$ zone-cycles/s at $168^3$ cells/GPU which is $\sim 3$ times slower than the V100 GPU performance for this test. Thus overall  we find that the performance of \theCode on RTX A6000 GPU is 3-6 times slower than that of V100 GPU, depending on the test setup used. The reason for this difference is that the peak double precision performance (FP64) of NVIDIA V100 GPU is 7.8 TeraFlop/s while that of NVIDIA RTX A6000 GPU is only 1.25 TeraFlop/s which makes A6000 $\sim$6 times slower than V100 theoretically. The corresponding costs are today $\sim$\$10,500 for a V100 while $\sim$\$4,500 for an RTX A6000, which makes a V100 GPU also more cost-efficient compared to an A6000 GPU.

\subsection{Weak Scaling}

Weak scaling is defined as how the performance of the code changes when the number of GPUs is increased, keeping the problem size per GPU fixed. This means that we increase the number of GPUs, and proportionately, increase the overall problem size to test how the time to solve the problem changes. In the ideal scenario, the time to solution should remain the same when we increase the number of GPUs, but in reality the time to solution generally increases on scaling up due to the communication overhead. We perform the weak scaling tests for \texttt{setupB} and \texttt{setupC} because they represent the setups that we'll run for production simulations. 

\begin{figure}
 	\includegraphics[width=0.7\columnwidth]{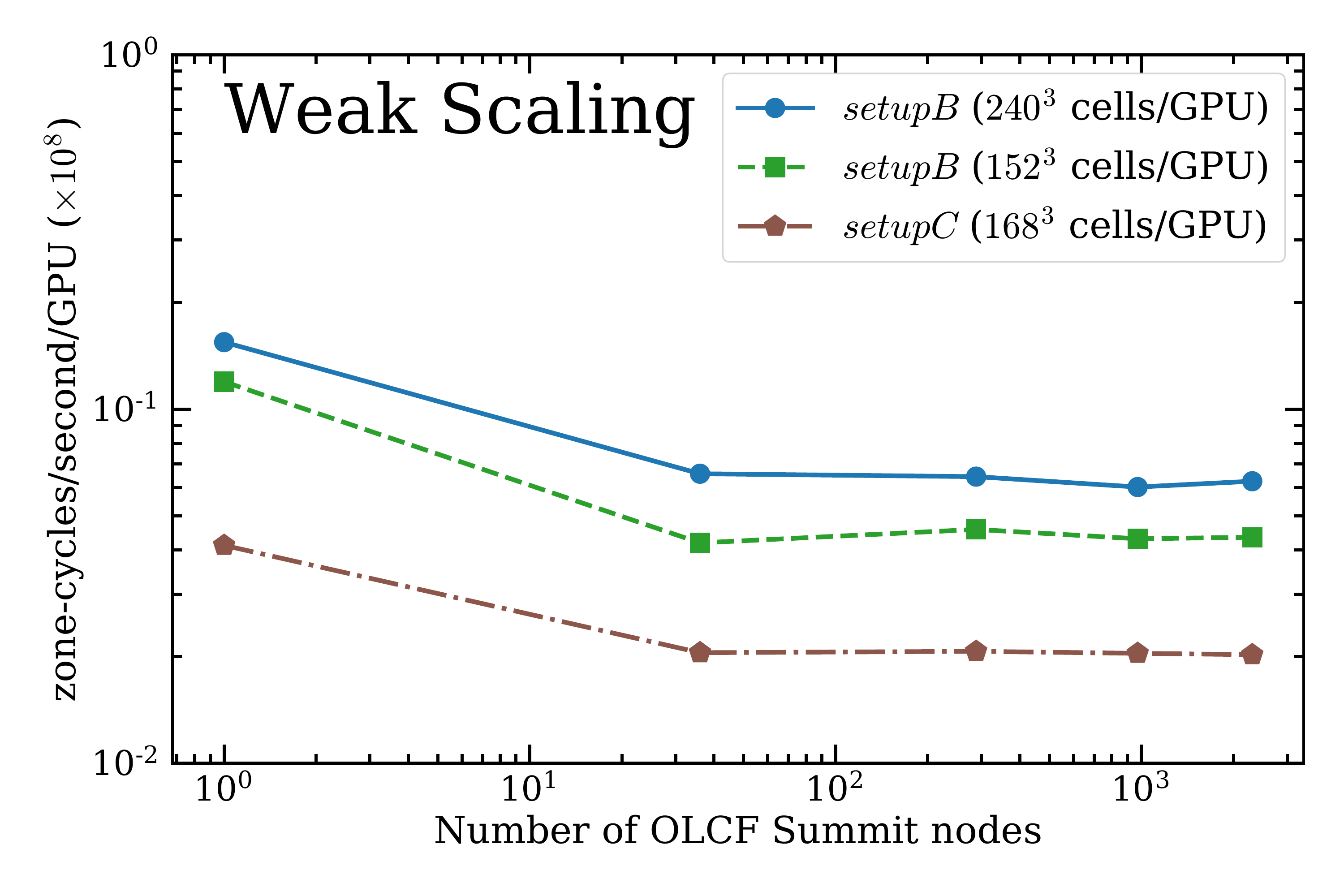}
    \caption{\theCode shows a weak scaling efficiency of $\sim40$-$50\%$ on 2304 nodes (13824 NVIDIA V100 GPUs)  with respect to single node performance on OLCF's supercomputer Summit. The weak scaling efficiency shows a steep drop going from 1 node to 36 nodes but remains almost constant going from 36 nodes up until 2304 nodes. We perform weak scaling test for \texttt{setupB} (static spacetime + Tabulated EoS + RK4) at $240^3$ and $152^3$ cells per GPU, and for \texttt{setupC} (\textit{dynamic} spacetime + Tabulated EoS + RK4) at $168^3$ cells per GPU. \texttt{setupB} shows better performance and scaling at $240^3$ cells/GPU compared to $152^3$ cells/GPU. \texttt{setupC} is the slowest due to extra computations for spacetime evolution but it shows the best scaling due to higher work load per GPU than other cases. } 
    \label{fig:weak_scaling}
\end{figure}

\begin{table}[]
\caption{Weak scaling grid setup and performance results for \texttt{setupB} and \texttt{setupC}}
\label{tab:weak_scaling}
\begin{tabular}{|c|c|ccc|ccc|}
\hline
\multirow{2}{*}{Nodes} & \multirow{2}{*}{GPUs} & \multicolumn{3}{c|}{\texttt{setupB} ($240^3$cells/GPU)} & \multicolumn{3}{c|}{\texttt{setupC} ($168^3$cells/GPU)} \\ \cline{3-8} 
 &  & \multicolumn{1}{c|}{Grid} & \multicolumn{1}{c|}{zone-cycles/s/GPU} & Efficiency & \multicolumn{1}{c|}{Grid} & \multicolumn{1}{c|}{zone-cycles/s/GPU} & Efficiency \\ \hline
1 & 6 & \multicolumn{1}{c|}{$720 \times 480 \times 240$} & \multicolumn{1}{c|}{$0.154 \times 10^8$} & 1.00 & \multicolumn{1}{c|}{$504 \times 336 \times 168$} & \multicolumn{1}{c|}{$0.0412\times 10^8$} & 1.00 \\ \hline
36 & 216 & \multicolumn{1}{c|}{$1440 \times 1440 \times 1440$} & \multicolumn{1}{c|}{$0.066 \times 10^8$} & 0.43 & \multicolumn{1}{c|}{$1008 \times 1008 \times 1008$} & \multicolumn{1}{c|}{$0.0205 \times 10^8$} & 0.50 \\ \hline
288 & 1728 & \multicolumn{1}{c|}{$2880 \times  2880 \times  2880 $} & \multicolumn{1}{c|}{$0.064 \times 10^8$} & 0.42 & \multicolumn{1}{c|}{$2016 \times  2016 \times  2016 $} & \multicolumn{1}{c|}{$0.0207 \times 10^8$} & 0.50 \\ \hline
972 & 5832 & \multicolumn{1}{c|}{$4320  \times  4320 \times  4320  $} & \multicolumn{1}{c|}{$0.060 \times 10^8$} & 0.39 & \multicolumn{1}{c|}{$3024  \times  3024 \times  3024  $} & \multicolumn{1}{c|}{$0.0204 \times 10^8$} & 0.50 \\ \hline
2304 & 13824 & \multicolumn{1}{c|}{$5760  \times  5760 \times  5760  $} & \multicolumn{1}{c|}{$0.063 \times 10^8$} & 0.40 & \multicolumn{1}{c|}{$4032  \times  4032 \times  4032  $} & \multicolumn{1}{c|}{$0.0203 \times 10^8$} & 0.49 \\ \hline
\end{tabular}
\end{table}

We perform the weak scaling test starting at 1 Summit node (6 V100 GPUs) going up to 2304 nodes (13824 V100 GPUs) with $240^3$ cells/GPU for \texttt{setupB} and $168^3$ cells/GPU for \texttt{setupC}. Table \ref{tab:weak_scaling} shows the grid setup used in each case for different number of nodes as well as the corresponding zone-cycles/s/GPU. We plot the results in Fig.~\ref{fig:weak_scaling}. We find that the weak scaling efficiency shows a steep drop going from 1 node to 36 nodes but remains almost constant after that up until 2304 nodes. It should be noted that there is no inter-node communication when using 1 node and this might be the reason that this case is much faster. We get a weak scaling efficiency of $\sim40\%$ for \texttt{setupB} and $\sim50\%$ for \texttt{setupC} at 2304 nodes when compared with the respective single node benchmarks. We also find that the weak scaling efficiency depends on the way we choose the grid, with more asymmetric grid choice providing better scaling efficiency. In the extreme case where we choose the grid as $1,399,680 \times 240 \times 240$ on 972 nodes, we get a weak scaling efficiency of $\sim98\%$ with respect to single node performance. This happens because when the boxes are stacked in a cubical shape, the amount of ghost cell communication increases as $\sim n_{\rm{boxes}}^3$ while when the boxed are stacked next to one another in an elongated shape, the amount of ghost cell communication increases only as $\sim n_{\rm{boxes}}$, where $n_{\rm{boxes}}$ is the number of boxes in the domain. Therefore in the general case of an asymmetric grid, the number of ghost cells decreases thus leading to less communication overhead and better scaling. 

Hence our weak scaling efficiency is limited to $\sim 40$-$50\%$ due to the communication required to fill ghost cells which, in turn, is limited by communication bandwidth (not latency). The solution to this problem is to either decrease the bandwidth requirements of our algorithms, increase available communication bandwidth, or possibly increase the overlap of computation and communication. The first can be achieved by using for example, Discontinuous Galerkin methods~\cite{Kidder:2017}, which reduce the number of ghost zones needed thus reducing bandwidth requirements. We have not yet attempted to implement such methods. The second depends on the hardware on which we run the code and we cannot change it. The latter depends on our code infrastructure and there is room for improvement in this case. In the current implementation of \carpetx, the next GPU kernel is only launched once all the ghost zones from the previous kernel have been filled. However, if the next kernel doesn't need the values from the previous kernel, it can be launched already while ghost zones are getting filled. Another strategy would be to start the computations of those cells of the next kernel which do not need ghost cells while ghost cells from previous kernel are getting filled, and then compute the remaining cells thereafter. This would lead to more overlap of computation and communication, and could dramatically improve the scaling efficiency of the code. \carpetx is still in active development, and work is being carried out to increase the computation-communication overlap.  

\subsection{Strong Scaling}

Strong scaling is defined as how the time to solution changes when number of GPUs is increased keeping the problem size fixed. In order to use GPUs efficiently, it is important to maximize the compute load per GPU which reduces the effect of communication overhead. Thus, strong scaling is not suited for GPUs because keeping the problem size fixed and increasing the number of GPUs decreases the work load per GPU which reduces the efficiency. However, it is also true that running the code at the most cost efficient setup is generally not the fastest. It is, therefore,  important in case of an actual production simulation to identify the optimal compute load per GPU. Hence we perform a strong scaling test by varying the compute load per GPU from $48^3$ cells/GPU to $240^3$ cells/GPU. 

\begin{figure}
 	\includegraphics[width=0.7\columnwidth]{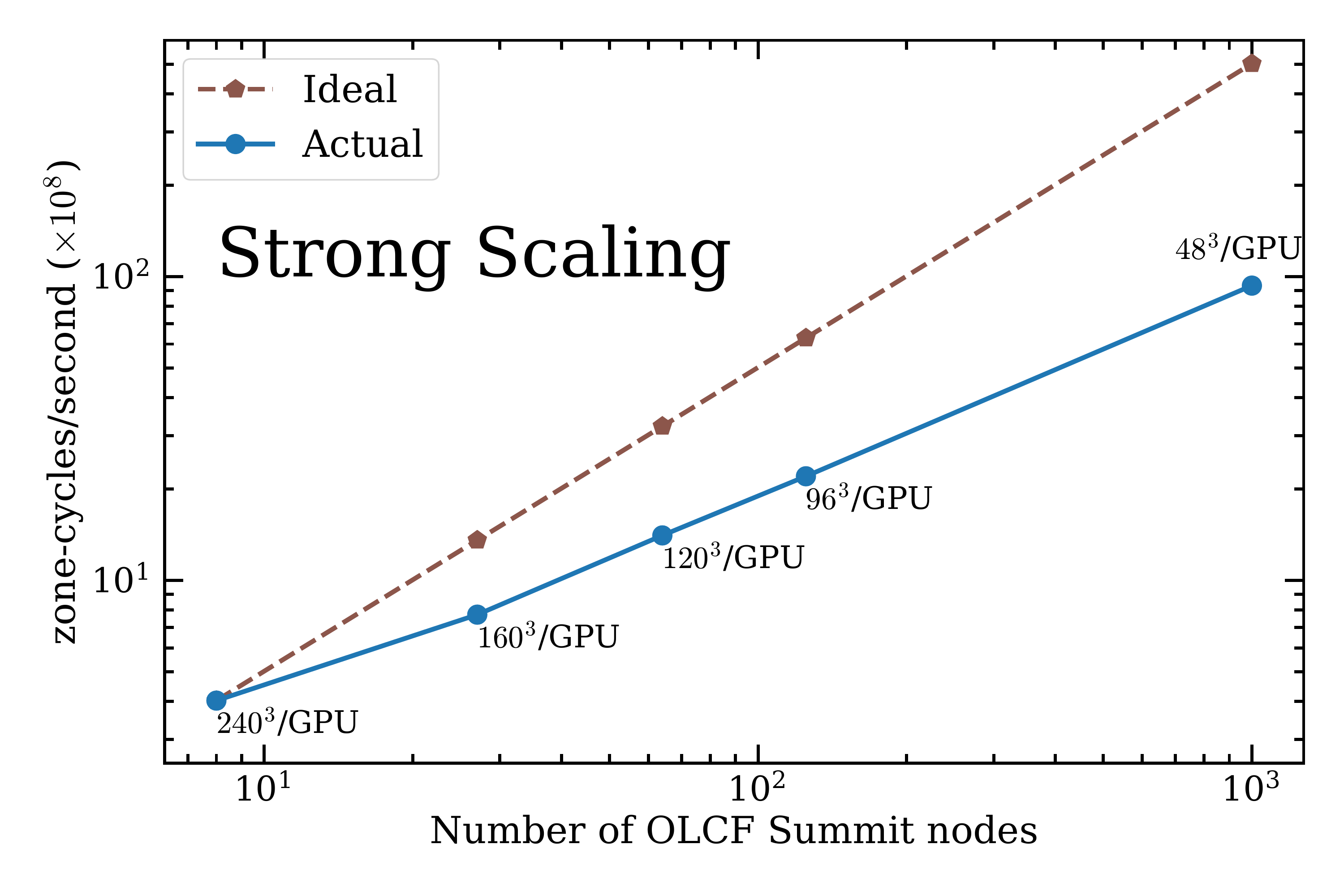}
    \caption{We perform the strong scaling test on \texttt{setupB} (static spacetime + Tabulated EoS + RK4)  by varying the number of nodes from 8 to 1,000 for a fixed problem size of  $1,440\times960\times480$. This increase in the number of nodes for the fixed problem size progressively reduces the compute load per GPU from $240^3$ cells/GPU (8 nodes) to $48^3$ cells/GPU (1,000 nodes). Lowering the compute load until $120^3$ cells/GPU (64 nodes) gives us a strong scaling efficiency of $\sim 44\%$ which is mainly because our weak scaling efficiency also drops up until $\sim 36$ nodes. However, we find that strong scaling efficiency drops further to $\sim 19\%$ going from 64 to 1,000 nodes even though our weak scaling efficiency didn't show any drop in this region. This is because going below $\sim120^3$-$160^3$ cells/GPU ($>27$-$64$ nodes), the compute load per GPU is not enough to keep all GPU threads occupied resulting in inefficient GPU utilization. } 
    \label{fig:strong_scaling}
\end{figure}

We perform the strong scaling test for \texttt{setupB} with a grid size of $1440 \times 960 \times 480$ which is equal to a total of $\sim872^3$ cells. We need to run this on a minimum of 8 Summit nodes (48 V100 GPUs) due to GPU memory constraints which corresponds to a compute load of $240^3$ cells/GPU. We then increase the number of nodes keeping the grid size fixed, going up to 1,000 nodes (6,000 V100 GPUs) which corresponds to a compute load of $48^3$ cells/GPU. Using the benchmarks, we calculate the total time (in hours) it will take to run this simulation till $10^5$ iterations and how much it will cost in total (in GPU-hours) for each case. We tabulate the results in Table \ref{tab:strong_scaling} and show the plot in Fig.~\ref{fig:strong_scaling}. We find that running this simulation on 8 nodes will take $\sim 46$ hours but cost only $\sim 2,200$ GPU-hours. On the other hand, running the same simulation on 1,000 nodes will take only $\sim 2$ hours but have a much larger cost of $\sim 12,000$ GPU-hours. The value of zone-cycle/s/GPU drops below $50\%$ at compute load of $120^3$ cells/GPU compared to $240^3$ cells/GPU which happens because we are increasing the communication overhead and decreasing the compute load per GPU both of which bring down the efficiency. So running on anything below a compute load of $\sim 120^3$ cells/GPU will be very inefficient. Running with $\sim 160^3$ cells/GPU is a good compute load for reasonable total runtime and cost efficiency on unigrid for \texttt{setupB}. This value will be different when we consider problems involving AMR or a different physics setup but similar reasoning can be applied there to arrive at a reasonable value of cells/GPU.  

\begin{table}[]
\caption{Strong scaling test for \texttt{setupB}  with a fixed problem size of $1440\times960\times480$ varying number of nodes from 8 to 1000 }
\label{tab:strong_scaling}
\begin{tabular}{|c|c|c|c|c|c|c|c|}
\hline
Nodes & GPUs & cells/GPU & zone-cycles/s & zone-cycles/s/GPU &  Efficiency & Time for $10^5$ iterations & Cost for $10^5$ iterations \\  
&    &    &    &    &     & (hours) & (GPU-hours) \\  \hline
8 & 48 & $240^3$/GPU & $4.02\times10^8$ & $0.084\times10^8$ & 1.00 & 45.8 & 2200 \\ \hline
27 & 162 & $160^3$/GPU & $7.72\times10^8$ & $0.048\times10^8$ & 0.57 & 23.9 & 3870 \\ \hline
64 & 384 & $120^3$/GPU & $14.07\times10^8$ & $0.037\times10^8$ & 0.44 & 13.1 & 5031 \\ \hline
125 & 750 & $96^3$/GPU & $22.03\times10^8$ & $0.029\times10^8$ & 0.35 & 8.4 & 6276 \\ \hline
1000 & 6000 & $48^3$/GPU & $93.46\times10^8$ & $0.016\times10^8$ & 0.19 & 2.0 & 11833 \\ \hline
\end{tabular}
\end{table}

\section{Summary}
\label{sec:summ}
We present a new GPU-accelerated dynamical-spacetime GRMHD code \theCode (\textbf{G}eneral \textbf{R}elativistic \textbf{a}ccelerated \textbf{M}agnetohydrodynamics on AMRe\textbf{X}) which runs efficiently and scales well across thousands of GPUs. \theCode is built upon the new AMR driver for Einstein Toolkit \carpetx and extends the capability of the Einstein Toolkit to simulate relativistic astrophysical systems such as core-collapse supernovae (CCSN) and binary neutron-star mergers (BNS) to GPU-based exascale supercomputers.   

\theCode features 3D adaptive-mesh refinement (AMR) and support for both analytic as well as tabulated equations of state. The AMR and GPU functionality of \theCode is enabled by the new AMR driver \carpetx which, in turn, leverages \amrex, an AMR library developed as part of the Block-Structured AMR Co-Design Center in the US DOE's Exascale Computing Project (ECP). We evolve the equations of general relativity using the Z4c formalism and the equations of ideal MHD using the Valencia formulation. \theCode includes 2nd-order accurate TVD (Total Variation Diminishing) and 5th-order accurate WENO (Weighted Essentially Non-Oscillatory) reconstruction methods. The Riemann solver \theCode employs is the HLLE (Harten-Lax-van Leer-Einfeldt) solver which is an approximate solver that uses a two-wave approximation to compute the update terms across the discontinuity at the cell interface. We use 3D Newton-Raphson (3D-NR) as the primary method for conservative-to-primitive transformation and fall back to the method of Newman \& Hamlin (an effective 1D method) in cases when the 3D-NR doesn't converge or converges to unphysical values. 

We have written \theCode from scratch in C++ with the core routines such as reconstruction methods and Riemann solver adapted from the already well-established code \grhydro. In order to test the validity and accuracy of the code, we perform a series of tests on static spacetime which include 1D MHD shocktubes, 2D magnetic rotor, and a 2D cylindrical explosion. We also test the code in dynamical spacetime using the linear oscillations of a 3D TOV star and extract the modes of gravity including the fundamental mode ($F$) and the first two overtones ($H1$, $H2$). In all the tests, we find very good agreement with the analytic results (wherever available) as well as the results reported by other codes in literature for similar test setups. We also perform single GPU benchmarks and scaling tests on OLCF's Summit to test the performance of \theCode. We find that the weak scaling efficiency of \theCode is $\sim$40-50\% on up to 2304 Summit nodes (13824 V100 GPUs) with respect to single node performance. We also find that higher compute load per GPU leads to more efficient performance, with a compute load of $240^3$ and $168^3$ cells/GPU leading to the most efficient performance for our production simulations with static and dynamical spacetime respectively on summit V100 GPUs. 

\theCode, for the first time, enables us to perform dynamical-spacetime general-relativistic astrophysics simulations such as core-collapse supernovae and neutron-star mergers at a speed 10 times faster and computational cost 30 times lesser compared to traditional CPU-based codes. Adjusting for the typical allocations awarded for CPU vs GPU supercomputers, one can perform $\sim$5 times more simulations with GPU-based codes such as \theCode compared to traditional CPU-based codes at a fraction of total runtime. This number will be higher if one maximizes the compute load per GPU which will lead to longer runtimes but even more cost efficiency. This makes GPU-computing both cost-effective as well as more environmentally friendly. We are currently testing and implementing a moment-based neutrino transport (M1) scheme~\cite{Radice:2022} to be used with \theCode for future production simulations. We also plan to include more compute intensive routines such as more accurate Riemann solvers Roe~\cite{Roe:1981ar} and Marquina~\cite{Donat:1996cs, Aloy:1999ne}. We expect the performance to be more efficient and the scaling to be the same or even better with the addition of these more compute intensive routines. 

\section{Acknowledgements}
\label{sec:ack}
This research was supported by National Science Foundation Grants No. NSF-2004879, NSF-2103680, NSF-1550514, NSF-2004157, ACI-1238993. The authors thank Ann Almgren, Weiqun Zhang, Andy Nonoka, Don Willcox, and the entire \amrex developer team for helpful discussions designing the code. SS and PM thank Matthew Liska for helpful discussions regarding scalability. SS thanks LBL for hosting him during the summer internship 2022 where some of this work was carried out. This work has benefited from participation in the NERSC December 2021 GPU hackathon~\cite{hackathon:web} and the authors like to especially thank the team mentors during the event Ronnie Chatterjee and Vassilios Mewes and the event leaders Iris Chen and Kevin Gott. The simulations were carried out on OLCF's Summit under allocation AST154 and LSU's DeepBayou.

Research at Perimeter Institute is supported in part by the Government of Canada through the Department of Innovation, Science and Economic Development and by the Province of Ontario through the Ministry of Colleges and Universities.

\bibliographystyle{unsrt}
\bibliography{bibliography2/misc_references,bibliography2/einsteintoolkit,bibliography2/references,bibliography2/jet_references,bibliography2/bh_formation_references,bibliography2/gw_references,bibliography2/sn_theory_references,bibliography2/grb_references,bibliography2/nu_obs_references,bibliography2/methods_references,bibliography2/eos_references,bibliography2/NSNS_NSBH_references,bibliography2/stellarevolution_references,bibliography2/nucleosynthesis_references,bibliography2/gr_references,bibliography2/nu_interactions_references,bibliography2/sn_observation_references,bibliography2/populations_references,bibliography2/pns_cooling_references,bibliography2/spectral_photometric_modeling_references,bibliography2/cs_hpc_references,bibliography2/numrel_references,bibliography2/radiation_transport_references,bibliography2/mhd_references,bibliography2/ns_merger_obs_references,bibliography2/sgr_magnetar_references,bibliography2/galactic_center_and_frb_references,bibliography2/publicationsfullpm}

\end{document}